\documentclass[a4paper,12pt]{article}
\usepackage{geometry}
\usepackage[english]{babel}
\usepackage[utf8]{inputenc}
\usepackage{amsmath}
\usepackage{indentfirst}
\usepackage{graphicx}
\usepackage{float}
\usepackage[colorinlistoftodos]{todonotes}
\usepackage{bbding}
\usepackage{pifont}
\usepackage{wasysym}
\usepackage{amssymb}
\usepackage{caption}
\usepackage{fullpage} 
\usepackage{caption,subcaption}
\usepackage{titling}
\usepackage{setspace}
\usepackage{tabularx}

\usepackage[colorlinks=true,citecolor=blue,linkcolor=blue,urlcolor=black]{hyperref}
\usepackage{booktabs,array,dcolumn} 
\usepackage[font=normalsize,labelfont=bf]{caption}
\usepackage[authoryear,round]{natbib}

\newcommand{\wt}{\widetilde}
\def\ov{\overline}
\providecommand{\tabularnewline}{\\}
\floatstyle{ruled}
\newfloat{algorithm}{tbp}{loa}
\providecommand{\algorithmname}{Algorithm}
\floatname{algorithm}{\protect\algorithmname}
\newcommand{\subtitle}[1]{%
	\posttitle{%
		\par\end{center}
	\begin{center}\large#1\end{center}
	\vskip0.9em}%
}
\newcommand\blfootnote[1]{%
	\begingroup
	\renewcommand\thefootnote{}\footnote{#1}%
	\addtocounter{footnote}{-1}%
	\endgroup
}
\usepackage{helvet}

\begin{document}
\title{The Interaction Between Credit Constraints and Uncertainty Shocks\textsuperscript{$\dagger$} }
\author{Pratiti Chatterjee\textsuperscript{$\ddagger$}, David Gunawan\textsuperscript{$\star$}, and Robert Kohn\textsuperscript{$\star\star$}}
\date{April 2020}
\maketitle		

\begin{abstract}\small{	
Can uncertainty about credit availability trigger a slowdown in real activity? This question is answered by using a novel method to identify shocks to uncertainty in access to credit. Time-variation in uncertainty about credit availability is estimated using particle Markov Chain Monte Carlo. We extract shocks to time-varying credit uncertainty and decompose it into two parts: the first captures the \textit{``pure"} effect of a shock to the second moment; the second captures total effects of uncertainty including effects on the first moment.  Using state-dependent local projections, we find that the \textit{``pure"} effect by itself generates a sharp slowdown in real activity and the effects are largely countercyclical. We feed the estimated shocks into a \textit{flexible} price real business cycle model with a collateral constraint and show that when the collateral constraint binds, an uncertainty shock about credit access is recessionary leading to a simultaneous decline in consumption, investment, and output.
		
		}
\end{abstract}

\blfootnote{\scriptsize{\textsuperscript{$\dagger$}We would like to thank Eric Swanson for excellent comments and feedback. 
}\\
	\textsuperscript{$\ddagger$} 
	Level 4, West Lobby, School of Economics, University of New South Wales Business School -- Building E-12, Kensington Campus, UNSW Sydney -- 2052, \textit{Email:} {pratiti.chatterjee@unsw.edu.au}, \textit{Phone Number:} {(+61) 293852150}. Website: \url{http://www.pratitichatterjee.com}\\
	\textsuperscript{$\star$} 39C. 164, School of Mathematics and Applied Statistics (SMAS), University of Wollongong, Wollongong, 2522; Australian Center of Excellence for Mathematical and Statistical Frontiers (ACEMS); National Institute for Applied Statistics Research Australia (NIASRA); \textit{Email}: dgunawan@uow.edu.au. \textit{Phone Number:} {(+61) 424379015}. \\
		\textsuperscript{$\star\star$} 	Level 4, West Lobby, School of Economics, University of New South Wales Business School -- Building E-12, Kensington Campus, UNSW Sydney -- 2052, and ACEMS \textit{Email:} {r.kohn@unsw.edu.au}, \textit{Phone Number:} {(+61) 424802159}}
\begin{center}
\small{{JEL Classification Codes: C32, E32, E44} \\
		{Keywords: Bayesian inference, Credit Fluctuations, Financial Crises, MCMC, Recessions,  Particle Methods,  Stochastic Volatility, Uncertainty Shocks }}
\end{center}
\newpage
\section{Introduction}

The aftermath of the `Great Recession' witnessed a surge of interest in examining the importance of uncertainty in generating business cycle fluctuations,
with an early and important contribution by \citet{Bloom1}.
More recently, the COVID-19 pandemic triggered financial market uncertainty - both through changes in measures such as the VIX or uncertainty about access to liquidity.\footnote{The role of credit market uncertainty in the face of the COVID-19 pandemic has been explicitly addressed in speeches by policy makers across financial institutions across countries. Statements issued by: Board of Governors of the Federal Reserve System: \url{https://www.federalreserve.gov/newsevents/pressreleases/bcreg20200323a.htm}, Reserve Bank of Australia: \url{https://www.rba.gov.au/speeches/2020/sp-gov-2020-03-19.html}, {Reserve Bank of India:} \url{https://www.rbi.org.in/Scripts/bs_viewcontent.aspx?Id=3847}.} Concurrently, the role of credit markets in shaping business cycle dynamics has gained traction with many authors documenting a link between credit build up in periods of expansion and the subsequent crash in recessions; e.g., \citet*{MS}, \citet*{JST1}. 
Our article proposes a novel description of the link between credit markets and uncertainty shocks. We exploit the dynamics of credit expansion and contraction to explicitly pin down the role of uncertainty shocks in credit markets  to quantify how a change in the second moment transmits into the real economy. 

We begin by presenting a new stylized fact that stems from the time-variation in the volatility of credit expansions and contractions. These changes are interpreted as time-variations in access to credit and a stochastic volatility model is estimated to document that time-varying uncertainty about credit access is a robust empirical feature. 
Our approach to estimating the  stochastic volatility model is based on recent advances in Bayesian methodology for state space models. In particular, we use the correlated version of pseudo marginal Metropolis-Hasting (PMMH) proposed by \citet{Deligiannidis:2018} that requires far fewer particles at every iteration of the Markov chain Monte Carlo step than the standard PMMH.

The article makes several important contributions. First, it documents time-variation in the volatility of different measures of credit-growth and credit-availability. In particular, it shows that uncertainty about access to credit rises sharply during downturns. 

Second, we fit a time series model of these shocks and quantify their effects on macroeconomic variables using local projection impulse response functions. This is done by separating the identified shock into two components – one that captures the independent or `pure' effect of a change in the second moment and the other capturing additional first-moment effects that are associated with effects of an uncertainty shock to credit access. To quantify the impact of shocks to uncertainty about access to credit, the extracted shocks are used to construct impulse responses using the local projections method \citep{Jorda}. Impulse responses are constructed for two scenarios: one in a linear model with no distinction between recessions and expansions; the other in a state-dependent version of local projections that allows for differences between recessions and expansions. 

For both the linear and state-dependent local projections, we find that the effects of shocks to uncertainty are contractionary; however in recessions, the effects are significantly larger. 
During downturns, a one-standard deviation shock to uncertainty about credit access triggers a sharp decline in auto-sales, durable goods consumption, and investment. Household leverage declines and the credit spreads increase sharply as well. More broadly, the unemployment rate shows a significant increase a few periods after the shock hits and non-durable consumption declines. If the uncertainty shock embeds first-moment implications, these effects are even larger.

The third main result is that the estimated uncertainty process is fed to a flexible price real business cycle model with collateral constraints. 
We show that when the collateral constraint binds, an increase in uncertainty about credit access leads to a simultaneous decline in consumption, investment, output, and hours worked. Using properties of the pruned third-order solution we isolate the precautionary response from the endogenous transmission operating through the higher order interaction terms. In the model, the shock transmits itself  by increasing the wedge between labor demand and working capital. Although labor supply increases through the initial precautionary response, the response through the interaction channel capturing the role of the wedge generates the decline in hours in equilibrium. Thus, even in the absence of sticky prices, our proposed shock can generate  features in the model that align with the stylized facts characterizing the transmission of shocks to aggregate uncertainty. 

The first group of studies connected to our paper examine the role of uncertainty in generating business cycle fluctuations. \citet*{Bloom1}, \citet*{FVVGQRRU}, \citet*{FVVGQKRR}, \citet*{BasuBundick}, and, \citet*{Bloometal} among others, motivate the role of shocks to the second moment as a driver of business cycles. These studies quantify the effects of a mean-preserving spread on business cycle dynamics. Uncertainty in these papers, however, stems from the time-varying volatility in exogenous shocks to aggregate productivity, aggregate demand, fiscal policy or borrowing costs. One of the challenges in the literature is understanding the origin and propagation of shocks to uncertainty. 

Our work is also related to studies that examine the role of financial conditions in transmitting uncertainty shocks as well as being a potential source of uncertainty shocks. \citet*{CFAGZ} empirically demonstrate that uncertainty shocks can be an important source of business cycle fluctuations; however, the severity of the impact increases when these shocks are allowed to interact with financial frictions. Recently, \citet*{LMN} find that sharply higher macroeconomic uncertainty in recessions is often an endogenous response to output shocks, while uncertainty about financial markets is a likely source of output fluctuations. Instead of imposing identifying restrictions and shock based restrictions, we overcome these challenges by building the measure of uncertainty shock from  time-variation in the volatility of credit expansion and contraction. 

A third related strand of the literature examines the differences between financial and nonfinancial recessions. The importance of credit growth in shaping business cycles has been studied in detail by \citet*{JST1} who show that the patterns of credit growth can predict the type of recovery with periods of high credit growth being followed by recessions that are deeper and longer. When the real effects of uncertainty shocks to credit access are analyzed, we find that these shocks (independent of changes in the first moment) can exacerbate the depth and duration of a recession by amplifying the slowdown. Our article presents a  potential explanation for the heightened depth and duration of recessions following a credit buildup. 

In terms of technique and approach to modeling uncertainty, our paper is related to \citet*{FVVGQRRU} and \newline \citet*{FVVGQKRR}. Like these papers we use sequential Monte Carlo methods; however, one advantage of our method is that the implementation is more efficient. 

The remainder of paper is organized as follows. Section 2 describes the data, modeling and estimation techniques used to uncover time-variations in credit-access. Section 3 discusses the results of the estimation. Section 4 empirically quantifies the real effects of the extracted uncertainty shocks. Section 5 presents a theoretical model to outline the transmission mechanism. Section 6 presents a robustness check of our results. Section 7 concludes.

\section{Testing the existence of stochastic volatility for access to credit \label{Univariate SV specification}}
\begin{figure}[!htbp]
	\begin{center}	
		\scalebox{0.09}{
			$\begin{array}{c}
			\includegraphics{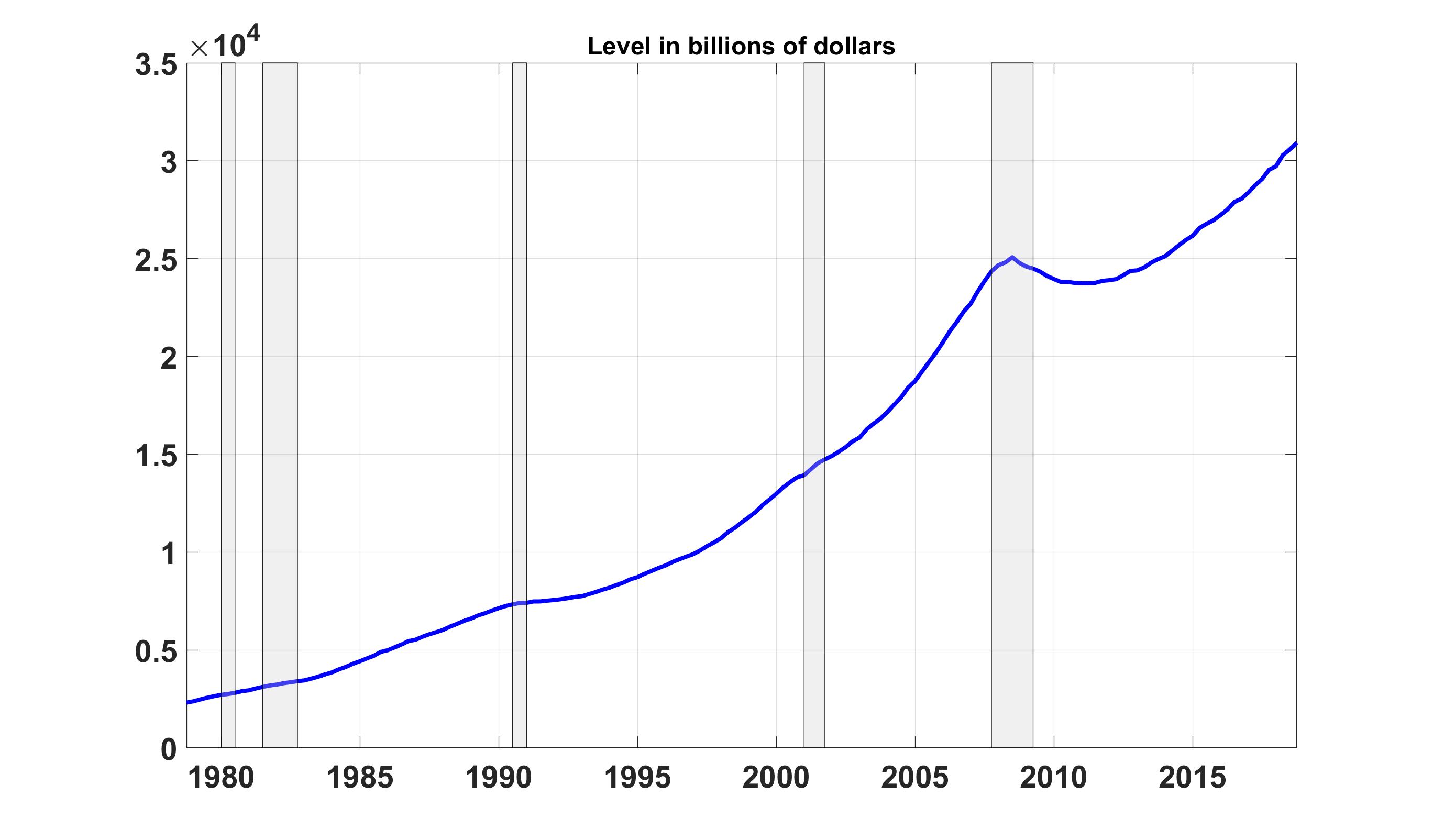} \\
			\includegraphics{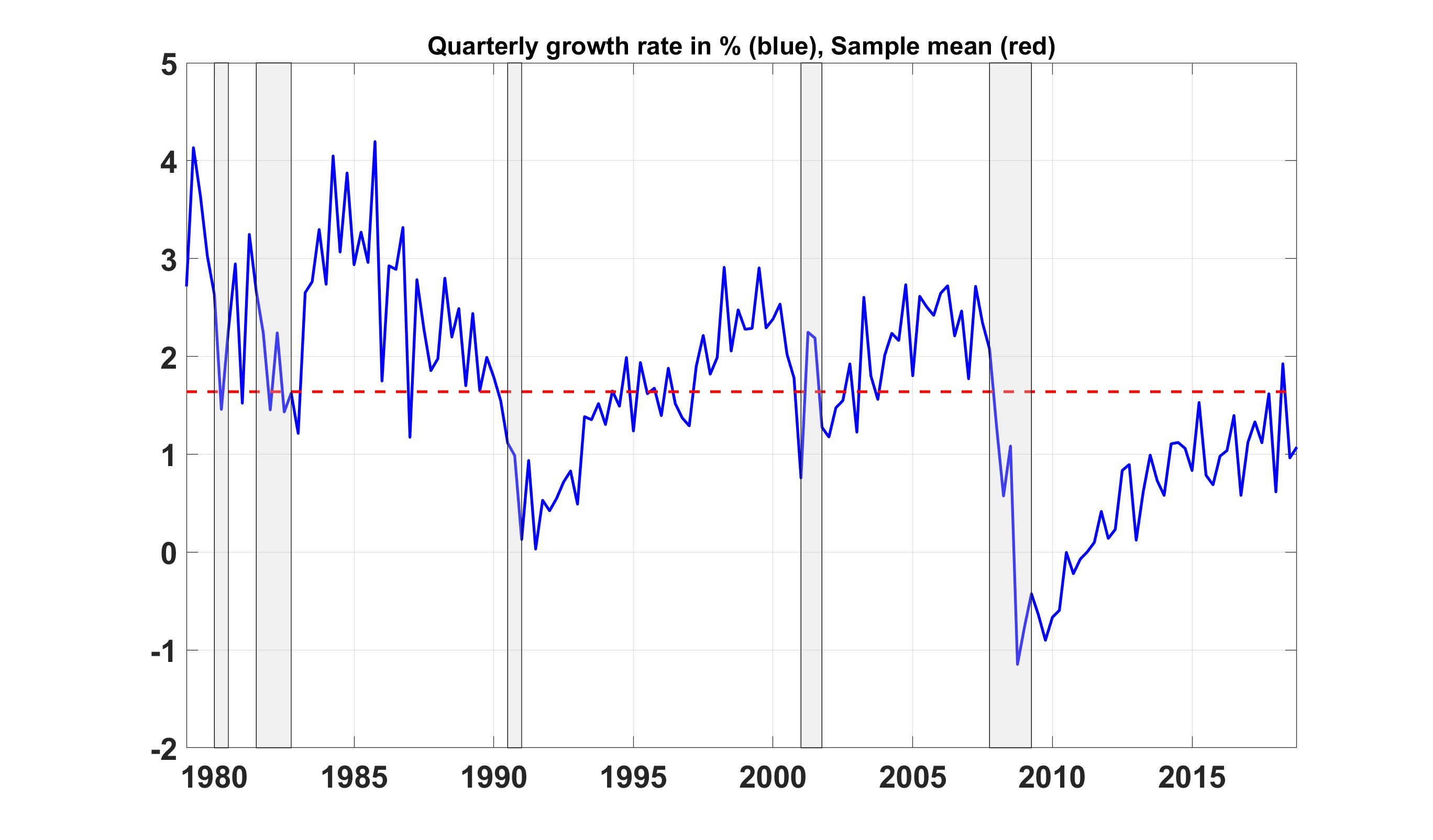} 			
			\end{array}$	}
	\end{center}
	\caption{Total credit extended to the nonfinancial sectors. Shaded areas indicate NBER recessions.}
	\label{fig:RawData}
\end{figure}

\paragraph{Data:} To examine the presence of time varying uncertainty in access to credit we begin by looking at the dynamics of the quarterly change in the growth rate of total credit extended to the nonfinancial sectors of the economy.\footnote{Credit is provided by domestic banks, all other sectors of the economy and non-residents. The "private nonfinancial sector" includes nonfinancial corporations (both private-owned and public-owned), households and non-profit institutions serving households as defined in the System of National Accounts 2008. The series have quarterly frequency and capture the outstanding amount of credit at the end of the reference quarter. In terms of financial instruments, credit covers loans and debt securities. Data has been valued using market valuation. For details see credit data on BIS statistics repository.} The estimation sample extends from 1978 Q1 to 2018 Q4; this series is denoted as $y_{t}$. Figure~\ref{fig:RawData} plots the quarterly level and growth rate of credit available to the nonfinancial sector. When the results in sections 3 and 4 are discussed, we will focus on uncertainty shocks to this particular measure of credit access. Section 6 carries out robustness checks accounting for different measures of credit access and show that the estimates characterizing the stochastic volatility in credit access is a robust empirical feature. 

\paragraph{Empirical Model:} The time variations in uncertainty about access to credit are isolated by first considering a stochastic volatility (SV) with leverage model for $y_t$, with observation equation 
\begin{eqnarray}
y_{t} & = & \phi_{y}y_{t-1}+\exp\left(h_{t}/2\right)\epsilon_{t},\;\;\epsilon_{t}\sim N\left(0,1\right),\;\;t=1,2,...,T;\label{observationequationbasic}
\end{eqnarray} 
$\epsilon_{t}$ is a standard normal random variable and $y_{0}=0$. The measure of credit growth is demeaned when estimating the model.  

The important feature of the model is that the log-volatility $h_{t}$ is not constant, 
but has a state transition equation that follows the AR(1) process, 
\begin{eqnarray}
h_{t} & = & \overline{\mu_h}+\phi_{h}\left(h_{t-1}-\overline{\mu_h}\right)+\tau\eta_{t},\;\;\eta_{t}\sim N\left(0,1\right),\;\;t=2,...,T;\label{statetransitionequationbasic}
\end{eqnarray}
\begin{eqnarray}
\left(\begin{array}{c}
\epsilon_{t}\\
\eta_{t}
\end{array}\right) & \sim & N\left(\left(\begin{array}{c}
0\\
0
\end{array}\right),\left(\begin{array}{cc}
1 & \rho\\
\rho & 1
\end{array}\right)\right)\label{eq:correlation};
\end{eqnarray}
$\eta_{t}$ is normally distributed with mean zero and unit variance. 
The persistence parameters $\phi_{y}$ and $\phi_{h}$ are restricted
to $\left(-1,1\right)$ to ensure that $\{y_t\}$ is a stationary process. The parameters $\overline{\mu_h}$ and $\tau$ control
the degree of mean volatility and stochastic volatility in $y_{t}$, respectively.
A high value of $\overline{\mu_h}$ implies a high mean volatility of $y_{t}$,
and a high value of $\tau$ implies a high degree of stochastic volatility.
The process $\{y_{t}\}$ is hit by both  $\epsilon_{t}$ and $\eta_{t}$;
the innovation $\epsilon_{t}$ to the observation $y_t$ changes the level of $y_{t}$; 
the innovation $\eta_{t}$ to the volatility of $y_t$ affects the standard deviation
of $\epsilon_{t}$. The parameter $\rho$ controls the strength of the dependence between
$\epsilon_{t}$ and $\eta_{t}$, and the size of the ``leverage effect" of the level shock $\epsilon_{t}$ on the volatility shock $\eta_{t}$. Therefore, uncertainty shocks are captured by $\eta_{t}$ (the disturbance of the latent volatility process). 
Equation \eqref{statetransitionequationbasic} can  be re-written as 
\begin{eqnarray}
h_{t}=\overline{\mu_{h}}+\phi_{h}\left(h_{t-1}-\overline{\mu_{h}}\right)+\tau\eta_{t}\label{eq:SVleveragealternative}, 
\end{eqnarray}
where $\eta_{t} = \rho\epsilon_{t-1}+\sqrt{1-\rho^{2}}\eta_{t}^{*}$, $\epsilon_{t-1}=\exp\left(-{h_{t-1}}/{2}\right)\left(y_{t-1}-\phi_{y}y_{t-2}\right)$
and $\eta_{t}^{*}\sim N\left(0,1\right)$. The standard SV model corresponds to $\rho=0$. 
The innovation $\eta_{t}$  captures the total effects of uncertainty shocks after taking into account the change in the first moment of the credit availability and the disturbance $\eta_{t}^{*}$ captures the (pure) independent effects of change in the second moment. 
For Bayesian inference, the joint posterior distribution of $\theta$ and $h_{1:T}$ is
\begin{equation}
p\left(\theta,h_{1:T}|y_{1:T}\right)=\frac{p\left(h_{1:T},y_{1:T}|\theta\right)p\left(\theta\right)}{p\left(y_{1:T}\right)};
\end{equation}
\begin{eqnarray*}
p\left(h_{1:T},y_{1:T}|\theta\right)=p\left(h_{1}|\theta\right)p\left(y_{1}|h_{1},\theta\right)p\left(h_{2}|h_{1},y_{1},\theta\right)p\left(y_{2}|h_{2},y_{1},\theta\right)\\ \prod_{t=3}^{T}p\left(h_{t}|h_{t-1},y_{t-1},y_{t-2},\theta\right)p\left(y_{t}|h_{t},y_{t-1},\theta\right);
\end{eqnarray*}
$p\left(\theta\right)$ is the prior density for $\theta$; and
\begin{eqnarray}
p\left(y_{1:T}\right)=\int_{\Theta}p\left(y_{1:T}|h_{1:T},\theta\right)p\left(h_{1:T}|\theta\right)p\left(\theta\right)dh_{1:T}d\theta,
\end{eqnarray}
is the marginal likelihood. The vector of model parameters for the univariate SV model with leverage is \{$\overline{\mu_h}$, $\phi_{y}$, $\phi_{h}$, $\tau$, $\rho$\}. 

The likelihood function for the univariate SV model is computationally intractable because it is a high dimensional integral over the latent states $h_{1:T}$. \citet{Andrieu:2010} propose the pseudo marginal Metropolis-Hastings
(PMMH) method for Bayesian inference in state space models; PMMH carries out Markov chain Monte Carlo (MCMC) on an expanded space
using an unbiased estimate $\widehat{p}\left(y_{1:T}|\theta,u\right)$ of the likelihood $p\left(y_{1:T}|\theta\right)$, where $u$ is the set of random numbers used to construct the likelihood estimator. Given that the PMMH is currently at $\left(\theta,u\right)$, the PMMH
sampler accepts a proposal $\left(\theta^{'},u^{'}\right)$ with the
acceptance probability 
\begin{equation}
\min\left\{ 1,\frac{\widehat{p}\left(y_{1:T}|\theta^{'},u^{'}\right)p\left(\theta^{'}\right)}{\widehat{p}\left(y_{1:T}|\theta,u\right)p\left(\theta\right)}\frac{q\left(\theta|\theta^{'}\right)}{q\left(\theta^{'}|\theta\right)}\right\}.\label{MH_acceptanceprob}
\end{equation}
Algorithm \ref{alg:standard} in Appendix \ref{alg:The-particle-marginal Mh} outlines the standard
PMMH method. 

\citet{Pitt:2012} show that under idealised conditions the variance of the log of the estimated
likelihood should be around 1 for the optimal performance of the PMMH
method, and that the performance of the method deteriorates exponentially as the variance
of the log of the estimated likelihood increases beyond 1. However, a drawback of PMMH is that it is sensitive to the size of the variance of the log of the estimated likelihood, so that for many problems it is computationally demanding to ensure that variance of the log of the estimated likelihood is around 1; e.g. \citet{FVVGQRRU} use
the PMMH method with $2000$ particles to obtain the unbiased estimate
of the likelihood. \citet{Deligiannidis:2018} refined the
PMMH method by correlating the pseudo random numbers $u$ and $u^{'}$ used in estimating
the likelihood at the current and proposed values of the Markov chain
to reduce the variance of  $\log\widehat{p}\left(y_{1:T}|\theta^{'},u^{'}\right)-\log\widehat{p}\left(y_{1:T}|\theta,u\right)$
appearing in \eqref{MH_acceptanceprob}. The correlated
PMMH approach helps the chain to mix even if highly variable
estimates of the likelihood are used; this means that generally far fewer particles are required at every iteration of the MCMC than for the standard PMMH. Algorithm \ref{alg:correlated} in Appendix \ref{alg:The-correlated-particle-marginal Mh}
describes the correlated PMMH algorithm.
The backward simulation algorithm
in \citet{Godsill2004} is used to sample the latent log-volatility; Algorithm \ref{alg:The-correlated particle filter}
in the appendix \ref{sec:The-Particle-Filter} describes the correlated particle filter algorithm to obtain the unbiased estimate of the likelihood. 

We follow \citet{Kim:1998} and choose the prior for the persistence
parameters $\phi_{y}$ and $\phi_{h}$ as $\left(\phi+1\right)/2\sim\textrm{Beta}\left(a_{0}=20,b_{0}=1.5\right)$,
i.e. 
\begin{equation}
p\left(\phi\right)=\frac{1}{2B\left(a_{0},b_{0}\right)}\left(\frac{1+\phi}{2}\right)^{a_{0}-1}\left(\frac{1-\phi}{2}\right)^{b_{0}-1}.
\end{equation}
The prior for $\tau$ is the half-Cauchy distribution, i.e. $p\left(\tau\right)=(2{I\left(\tau>0\right)})/(\pi{(1+\tau^{2})})$
, the prior for $p\left(\overline{\mu_h}\right)\propto1$, and the prior for $p\left(\rho\right)\sim U\left(-1,1\right)$.
These prior densities cover most possible values in practice,
are non-informative, and independent. 

The correlated PMMH sampler was run for 15000 iterations, with the initial
5000 iterations discarded as burn-in. We use an adaptive random walk proposal for $q\left(\theta^{'}|\theta\right)$
and, following \citet{Garthwaite:2016}, adaptively tune the scaling factor of the covariance matrix. This
enables us to pre-specify the overall acceptance probability before
running the correlated PMMH method. In the examples, the overall
acceptance probability is set as 25\% and the number of particles to $100$. 


\section{Empirical Results}
Table \ref{tab:Posterior-Mean-Estimates SV} reports the posterior
mean estimates of the stochastic volatility parameters (with 95\% credible intervals
in brackets) for the growth rate in total credit available to the nonfinancial sector. The table shows that:
(1) the parameter estimates of the basic SV model assuming $\rho=0$ are similar to the parameter estimates of the SV model with leverage. Next, we consider the parameter estimates
of the SV model with leverage; (2) the average volatility of an innovation
to total credit, $\overline{\mu_h}$, is large revealing a large
degree of volatility in the total credit and the posterior of $\overline{\mu_h}$
is tightly concentrated; (3) there is substantial stochastic volatility in the total credit series (a large $\tau$);
(4) the shocks to the level and log-volatility of total credit
are quite persistent (large $\phi_{y}$ and $\phi_{h}$);  (5) the credible interval for the correlation parameter $\rho$ is 
wide. 
\begin{table}[H]
	\caption{Posterior Mean Estimates (with 95\% credible intervals in brackets)
		of the standard SV Model and the SV model with leverage. NA means that the parameter does not appear in the model. \label{tab:Posterior-Mean-Estimates SV}}
	
	\centering{}%
	\begin{tabular}{ccc}
		\hline 
		Parameters & SV & SV-leverage\tabularnewline
		\hline 
		$\overline{\mu_h}$ & $\underset{\left(-10.77,-9.32\right)}{-10.12}$ & $\underset{\left(-10.83,-9.61\right)}{-10.23}$\tabularnewline
		$\phi_{h}$ & $\underset{\left(0.64,0.99\right)}{0.89}$ & $\underset{\left(0.75,0.98\right)}{0.91}$\tabularnewline
		$\phi_{y}$ & $\underset{\left(0.73,0.91\right)}{0.83}$ & $\underset{\left(0.75,0.91\right)}{0.83}$\tabularnewline
		$\tau$ & $\underset{\left(0.09,0.66\right)}{0.29}$ & $\underset{\left(0.12,0.53\right)}{0.27}$\tabularnewline
		$\rho$ & NA & $\underset{\left(-0.16,0.95\right)}{0.50}$\tabularnewline\hline  
	\end{tabular}
\end{table}

\begin{figure}[H]
	\begin{center}	\scalebox{0.25}{
			$\begin{array}{c}
			\includegraphics{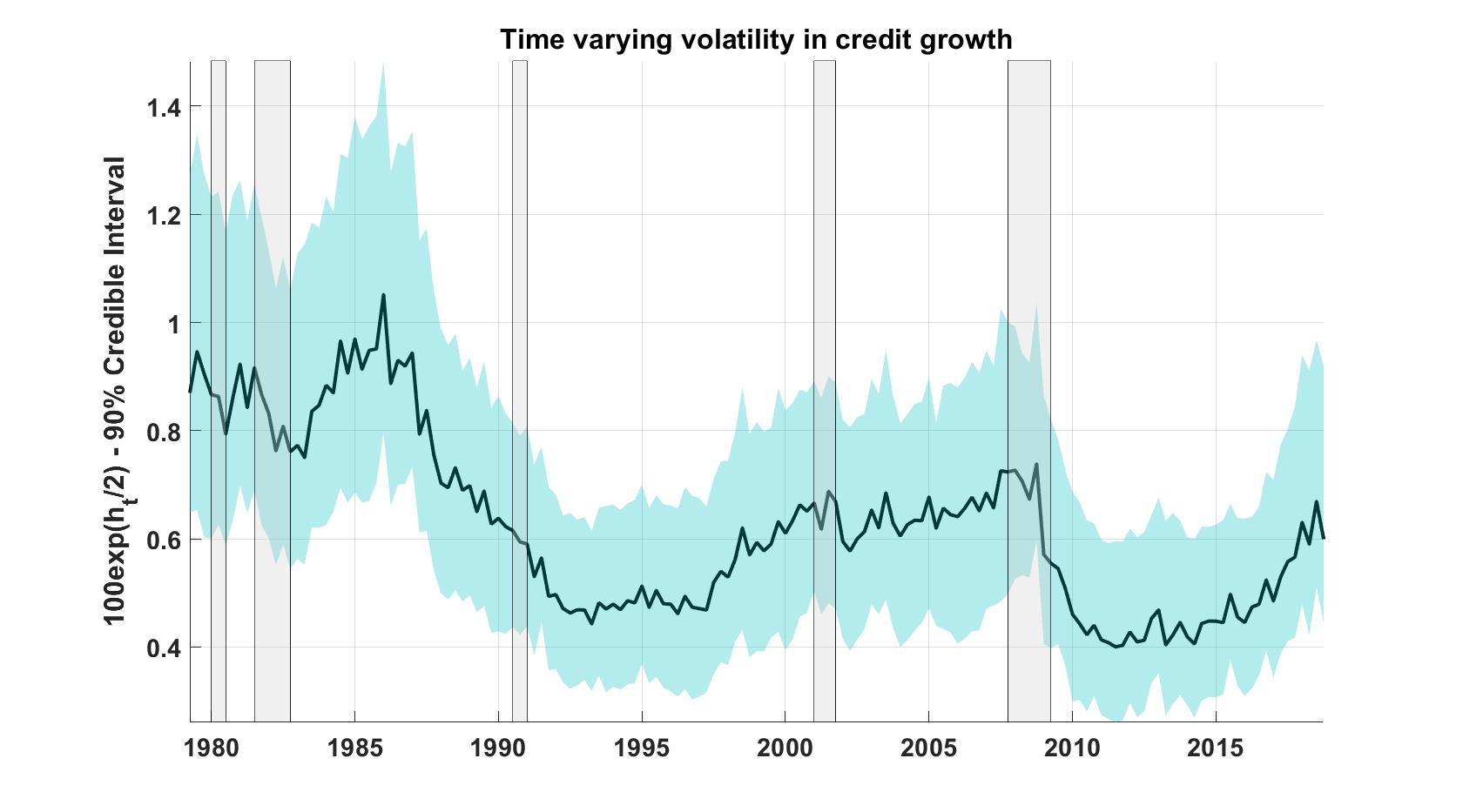} 
			\end{array}$	}
	\end{center}
	\caption{Time varying volatility in credit growth. We plot the posterior mean estimate of the process describing the standard deviation in credit-growth -- $100\exp(h_t/2)$. Shaded grey areas show NBER recessions.}
	\label{fig:EstShocks0}
\end{figure}

\begin{figure}[H]
	\begin{center}	\scalebox{0.25}{
			$\begin{array}{c}
			\includegraphics{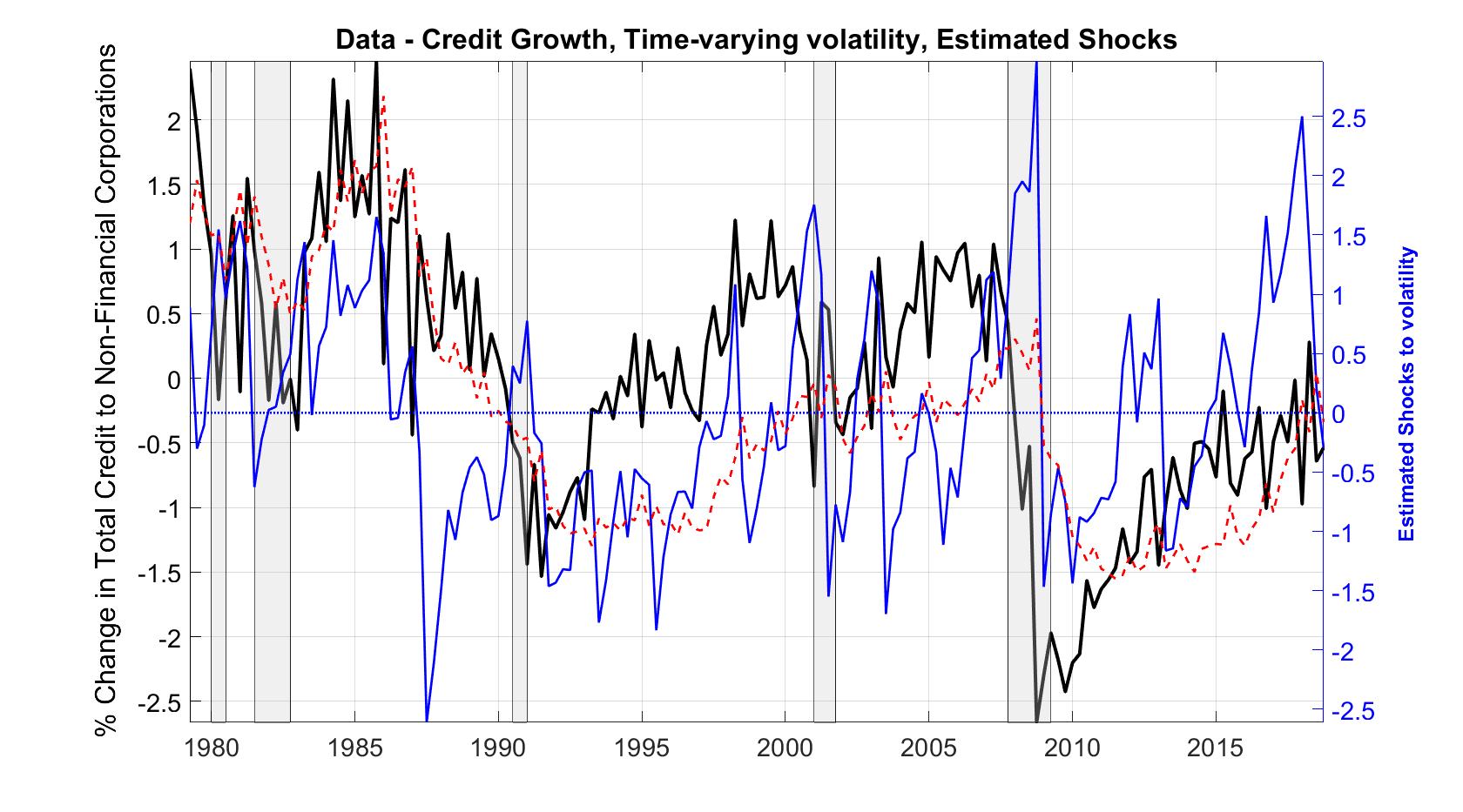} 
			\end{array}$	}
	\end{center}
	\caption{Estimated shocks to the volatility in credit-growth. We plot the posterior mean estimate of the shocks that capture independent changes in uncertainty in blue, Time-Varying Volatility in red, and Data on quarterly growth rate for total credit available to the nonfinancial sector in black. Shaded gray areas denote NBER recessions. We standardize the data on the growth rate of credit, extracted shocks and estimated volatility to capture the relative difference in the size of shocks between recessions and expansions.}
	\label{fig:EstShocks1}
\end{figure}

Figure~\ref{fig:EstShocks0} plots the posterior mean estimate for the volatility process and its 90\% credible interval; the figure shows that the second moment characterizing access to credit displays significant time variations. Our measure effectively distinguishes periods of high and low liquidity. The earlier part of the sample, between 1982 and 1986, is a period of high credit growth and high uncertainty in the credit growth. This time period also has banking deregulation. The turning point in the late eighties shows that there is a decline in the volatility as the process of deregulation becomes fully integrated. This decline in credit volatility continues into the late 1990s when the economy witnesses a dot-com boom and bust. Figure~\ref{fig:EstShocks1} plots the estimated posterior mean for the volatility, extracted shocks and the growth rate of credit and helps us understand how the extracted shock moves with the business cycle process. 
Table ~\ref{table:LeadLag} reports the correlation at various leads and lags for our measure of time-varying volatility and existing measures of uncertainty. While a  positive correlation is observed, it is important to note that the estimate does not move together with these measures. 
The shocks are extracted from credit market itself, so while measures such the VIX capture aggregate uncertainty they do not tell us whether it originates in the macro or the financial sectors of the economy i.e. is it spiking because of a demand shock or a financial shock.

\begin{table}[H]
\caption{Lead/Lag Correlation: $\textrm{Corr}\left(\sigma_{h,t},Var_{x,t+k}\right)$,
	where $\sigma_{h,t}=\exp\left({h_{t}}/{2}\right)$ is our measure,
	$Var_{x,t+k}$ are the alternative measures of uncertainty defined in column 1, and $k$ is the number of quarters
	ahead}
	\label{table:LeadLag}\scriptsize{
		\begin{center}
			\begin{tabular}{p{0.375\textwidth} p{0.01\textwidth}  p{0.035\textwidth}p{0.035\textwidth}
				p{0.035\textwidth}  p{0.035\textwidth}p{0.035\textwidth} p{0.035\textwidth}  p{0.035\textwidth}p{0.035\textwidth}}\hline\hline				
Alternative measures of uncertainty & k  & -3 & -2 & -1 & 0 & 1 & 2 & 3 & 4 \\	\hline
BBD -- Economic Policy Uncertainty Index &  & 0.21 & 0.23 & 0.24 & 0.29 & 0.27 & 0.25 & 0.27 & 0.31\\
JLN -- Macro Uncertainty Index/h=1
&  & 0.37 & 0.40 & 0.43 & 0.48 & 0.52 & 0.53 & 0.54 & 0.53\\
JLN -- Real Uncertainty Index/h=1 &  & 0.29 & 0.31 & 0.34 & 0.38 & 0.40 & 0.42 & 0.42 & 0.42\\
JLN -- Financial Uncertainty Index/h=1 & & 0.10 & 0.13 & 0.18 & 0.23 & 0.27 & 0.29 & 0.32 & 0.31\\
VIX &  & 0.19 & 0.22 & 0.24 & 0.31 & 0.35 & 0.37 & 0.42 & 0.44\\\hline			
			\end{tabular}
	\end{center}}
\end{table}
Thus far, the dynamics of the second moment characterizing access to credit are examined. This measure distinguishes periods of easy access to liquidity from periods of a liquidity crunch and low credit-access. Next, we consider the importance of this process in generating business cycle fluctuations.
\section{Real effects of uncertainty about  credit access}
Do shocks to uncertainty about credit access have real effects? To answer this question we use the extracted shocks to uncertainty about the credit growth rate and construct  local projections as in \citet{Jorda}. Equation~\eqref{eq:linearLP} describes our specification for constructing the regime-independent responses of macroeconomic variables at horizon $h$ for shocks to uncertainty about credit-access in period $t$,
\begin{equation}\label{eq:linearLP}
x_{t+h}=\alpha_h+\psi_h(L)z_{t-1}+\beta_h{shock_t}+\epsilon_{t+h}\text{ ;}
\end{equation}
$x$ is the macroeconomic variable of interest, $z$ is a vector of control variables, $\psi_h(L)$ is a second order polynomial in the lag operator, and shock is the identified shock -- either $\eta_t^*$ or $\eta_t$ from the stochastic volatility model in section 2. The baseline  control variables include lags of the GDP growth rate, lags of the dependent variable, lags of a financial stress index\footnote{We use the Chicago Fed’s National Financial Conditions Index (NFCI) to control for the state of financial conditions.}, and lags of  the quarterly credit growth rate. The macroeconomic variables considered for constructing local projections are credit-dependent measures of real activity (total vehicle sales and durable consumption) as well as broader measures of macroeconomic activity (aggregate consumption, non-durable consumption, investment and the unemployment rate).
To understand the effects on credit markets, we also examine the effect on the growth rate of household leverage and the credit spread between Baa and Aaa corporate bonds. All growth rates are computed for real variables. 

We follow \citet{RZ} to account for state dependence and examine the existence heterogeneity in the response of macroeconomic variables to uncertainty about credit access in recessions and expansions; and extend the linear model in Equation~\eqref{eq:linearLP} to,
\begin{multline}\label{eq:nonlinearLP}
x_{t+h}=I_{t-1}[\alpha_{R,h}+\psi_{R,h}(L)z_{t-1}+\beta_{R,h}{shock_t}+\epsilon_{t+h}]\\+(1-I_{t-1})[\alpha_{NR,h}+\psi_{NR,h}(L)z_{t-1}+\beta_{NR,h}{shock_t}+\epsilon_{t+h}] \text{ .}
\end{multline}	
Equation~\eqref{eq:nonlinearLP} allows for state dependence in calculating impulse responses; $I_{t-1}$ is a dummy variable indicating the state of the economy when the shock hits. All the model coefficients are allowed to vary according to the state of the economy. To account for the serial correlation in the error terms, the Newey-West correction is used for the standard errors.

The results are presented in two parts. First,  Figures~\ref{fig:MacroVarInd} and ~\ref{fig:CreditInd}, isolate the effects of a shock to the second moment \textit{only} by removing the effects of changes in the first moment. Second, Figures  ~\ref{fig:MacroVarCorr} and ~\ref{fig:CreditCorr} plot the total effect of a shock to the second moment characterizing access to credit after accounting for a change in the first moment of credit growth rate and its uncertainty. This decomposition provides a novel insight by capturing the pure effects of time-variation in uncertainty and how it interacts with changes in the first moment. The impulse responses are computed using growth rates (log-first differences) of total vehicle sales, aggregate consumption, durable consumption, non-durable consumption, investment, unemployment and household leverage and standardizing the shocks such that the coefficients in Equation~\eqref{eq:linearLP} and ~\eqref{eq:nonlinearLP} can be interpreted as the elasticity of the relevant variable with respect to a one-standard deviation change in uncertainty.
\subsection{Impact of a ``pure” uncertainty shock  \texorpdfstring{$\mathbf{\eta_t^*}$}{Lg}}
\begin{figure}[H]
	\begin{center}	\scalebox{0.14}{
			$\begin{array}{cc}
			\includegraphics{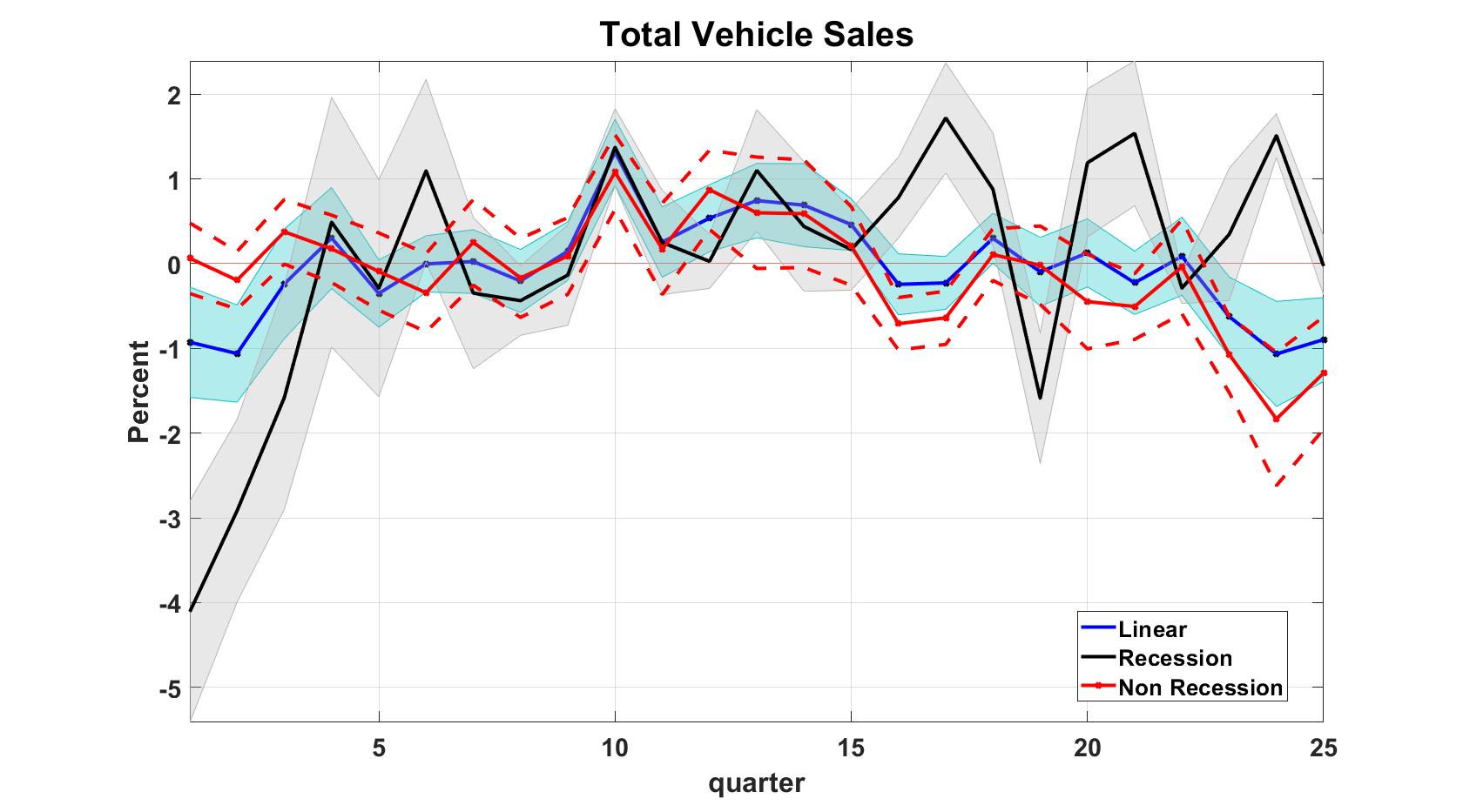} &
			\includegraphics{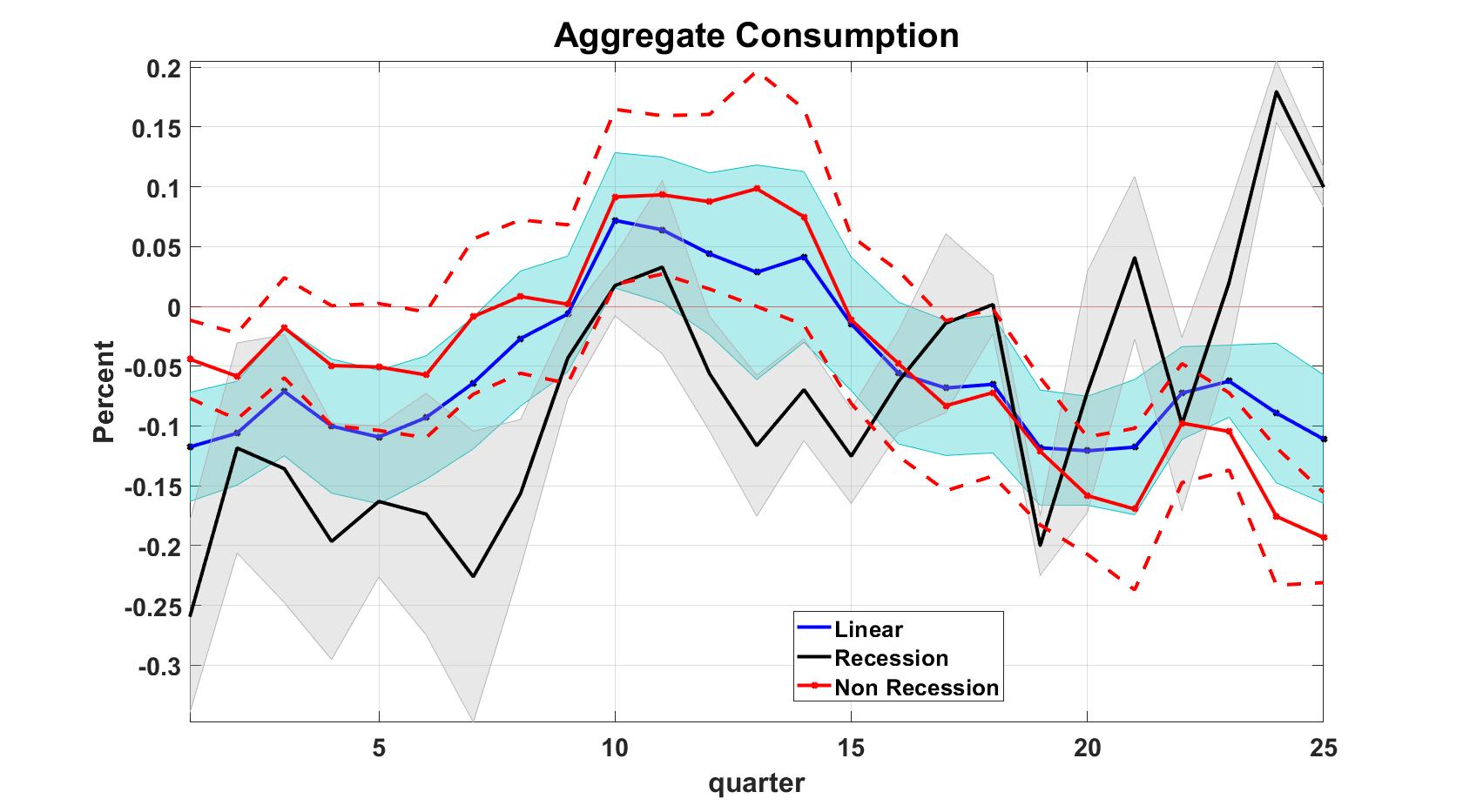} \\
			\includegraphics{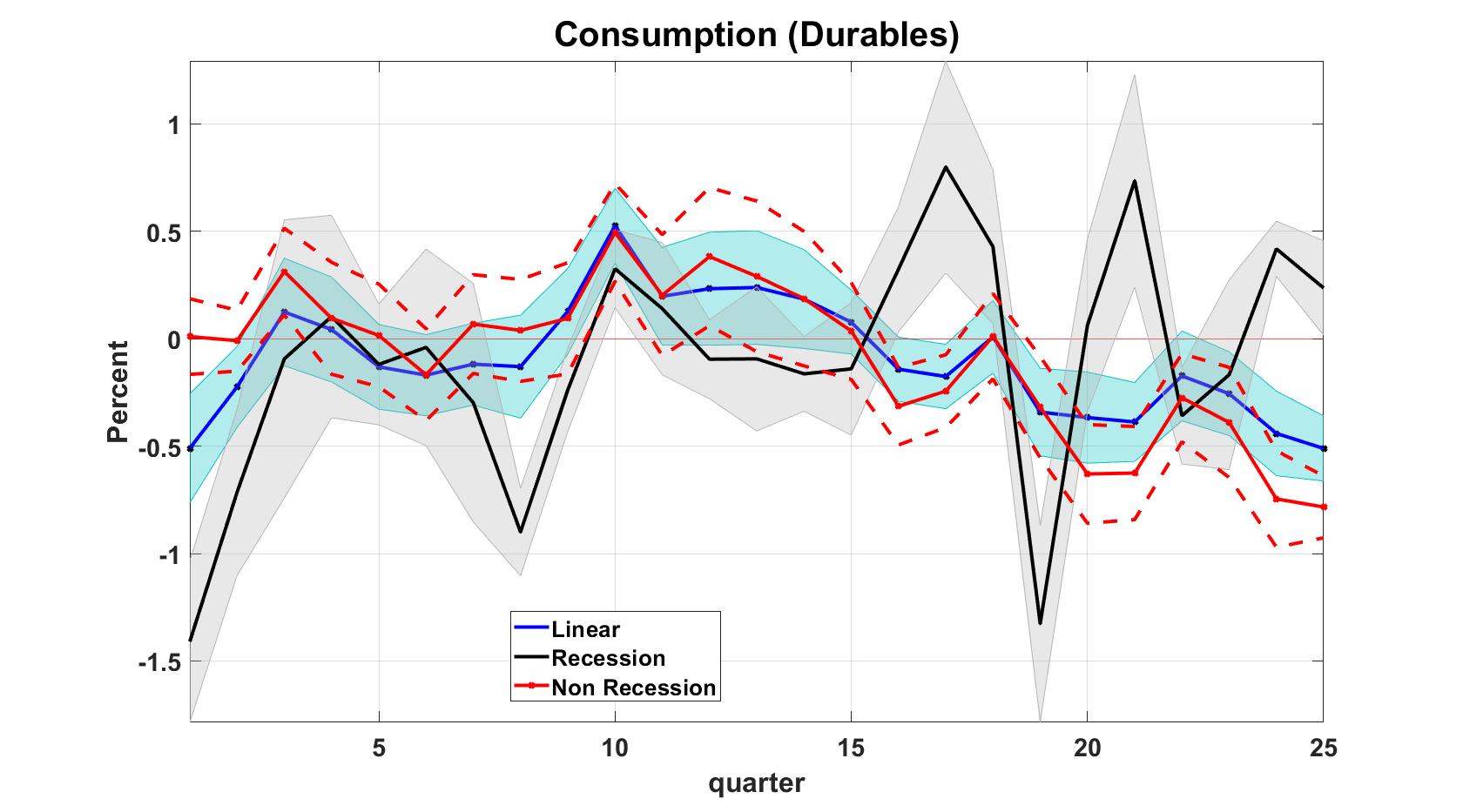} &
			\includegraphics{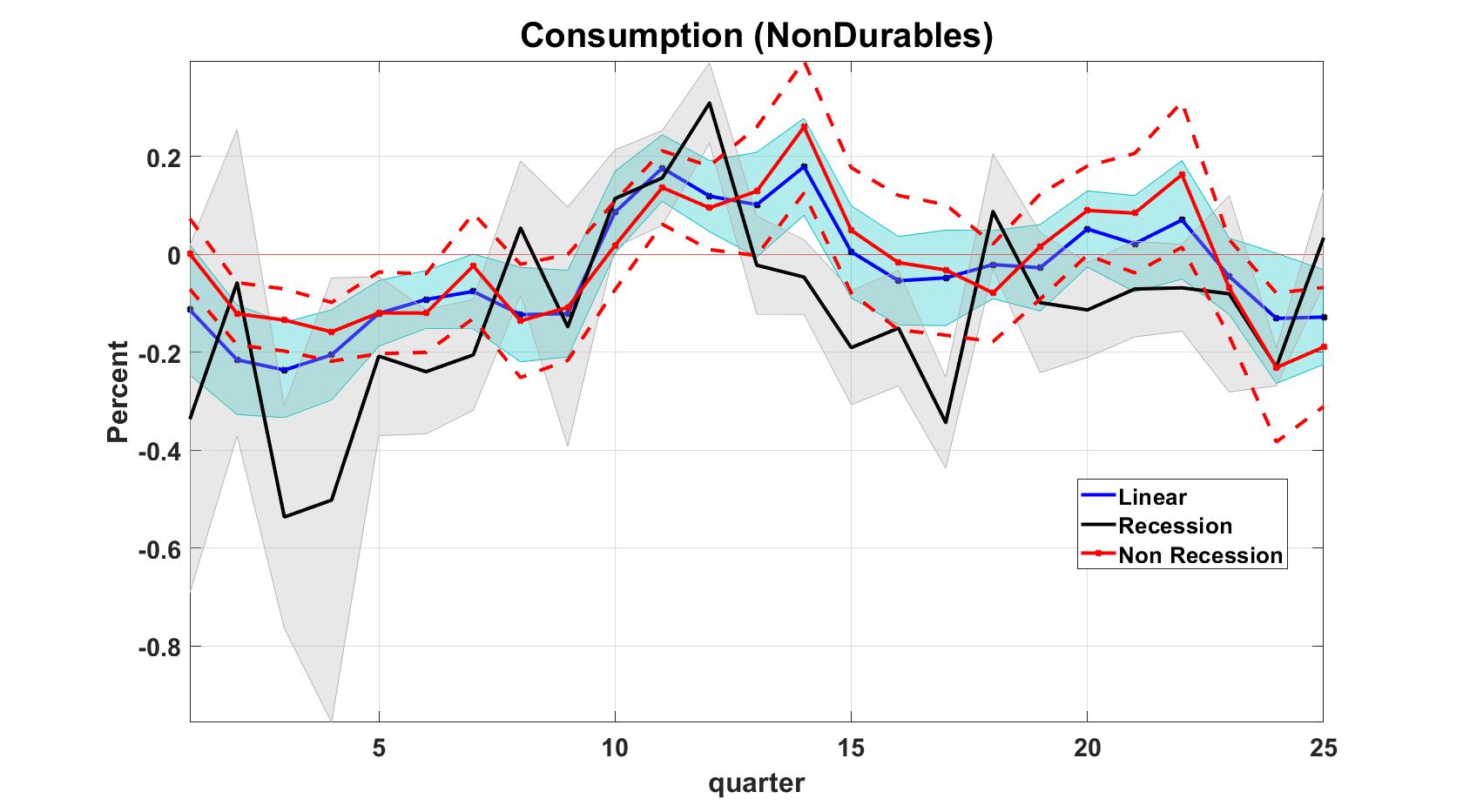} \\
			\includegraphics{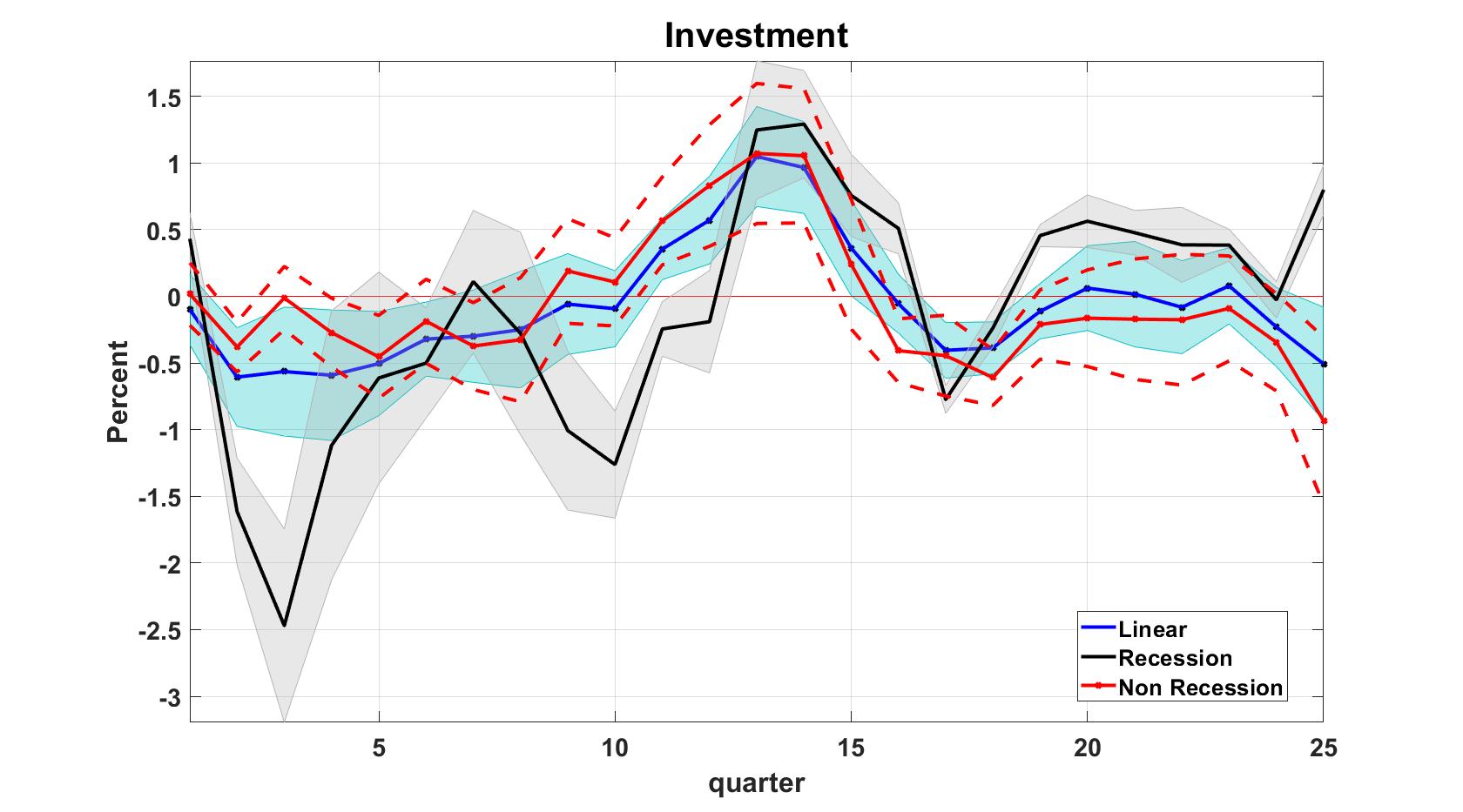}&
			\includegraphics{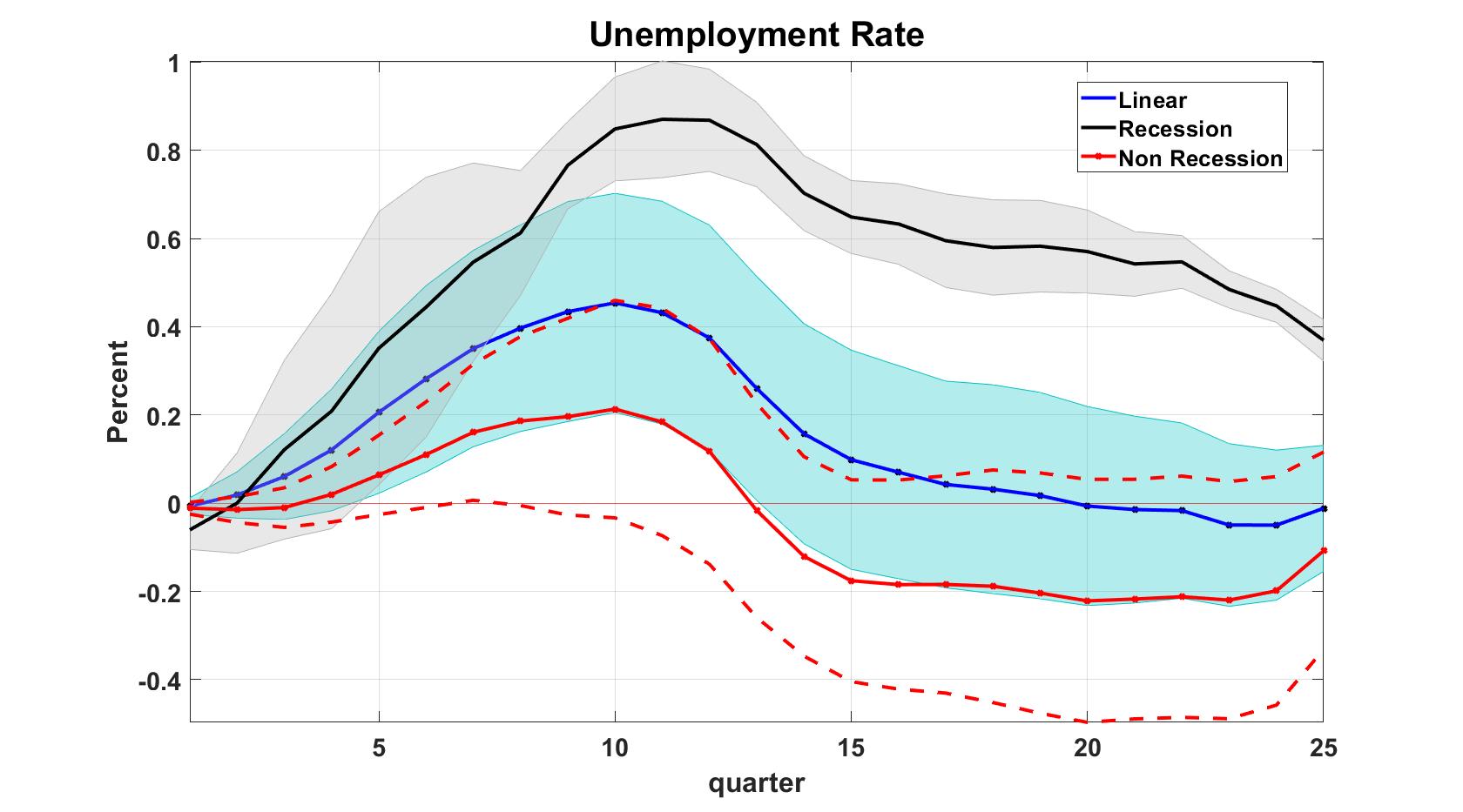}		\\	
			\end{array}$	}
	\end{center}
	\caption{Impact of changes in \texorpdfstring{${\eta_t^*}$}{Lg} on real activity. Impulse responses calculated for the quarterly real growth rate of total vehicle sales aggregate consumption, expenditure on durable consumption, expenditure on non-durable consumption, investment and the unemployment rate. The blue line is the effect of a one standard deviation shock to \texorpdfstring{${\eta_t^*}$}{Lg} in the linear model (shaded blue area -- 68\% CI). The black line is the effect of a one standard deviation shock to \texorpdfstring{${\eta_t^*}$}{Lg} in recessions (shaded gray area -- 68\% CI). The red line is the effect of a one standard deviation shock to \texorpdfstring{${\eta_t^*}$}{Lg} in expansions (dashed red line -- 68\% CI).}
	\label{fig:MacroVarInd}
\end{figure}

\begin{figure}[H]
	\begin{center}	\scalebox{0.14}{
			$\begin{array}{cc}
			\includegraphics{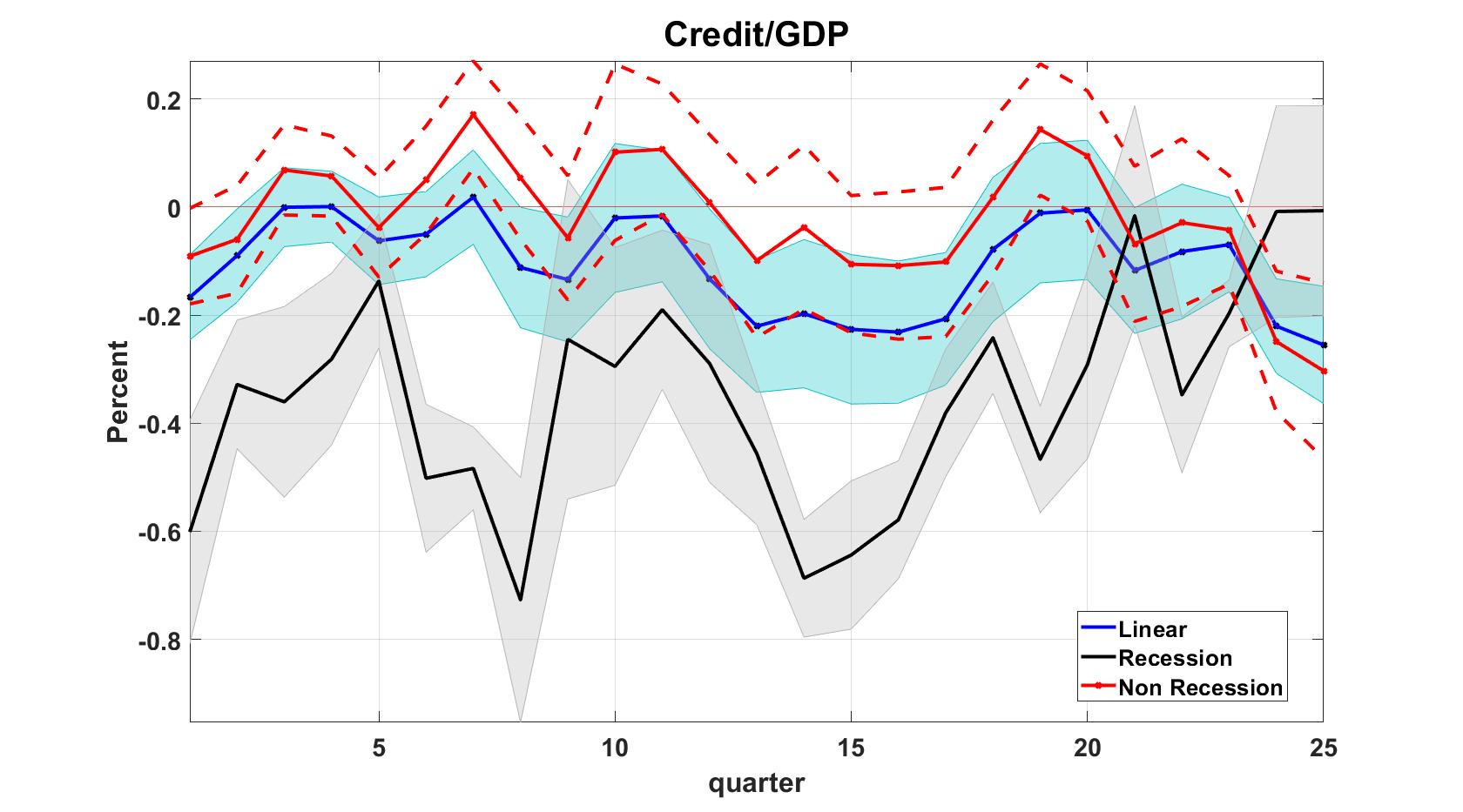} &
			\includegraphics{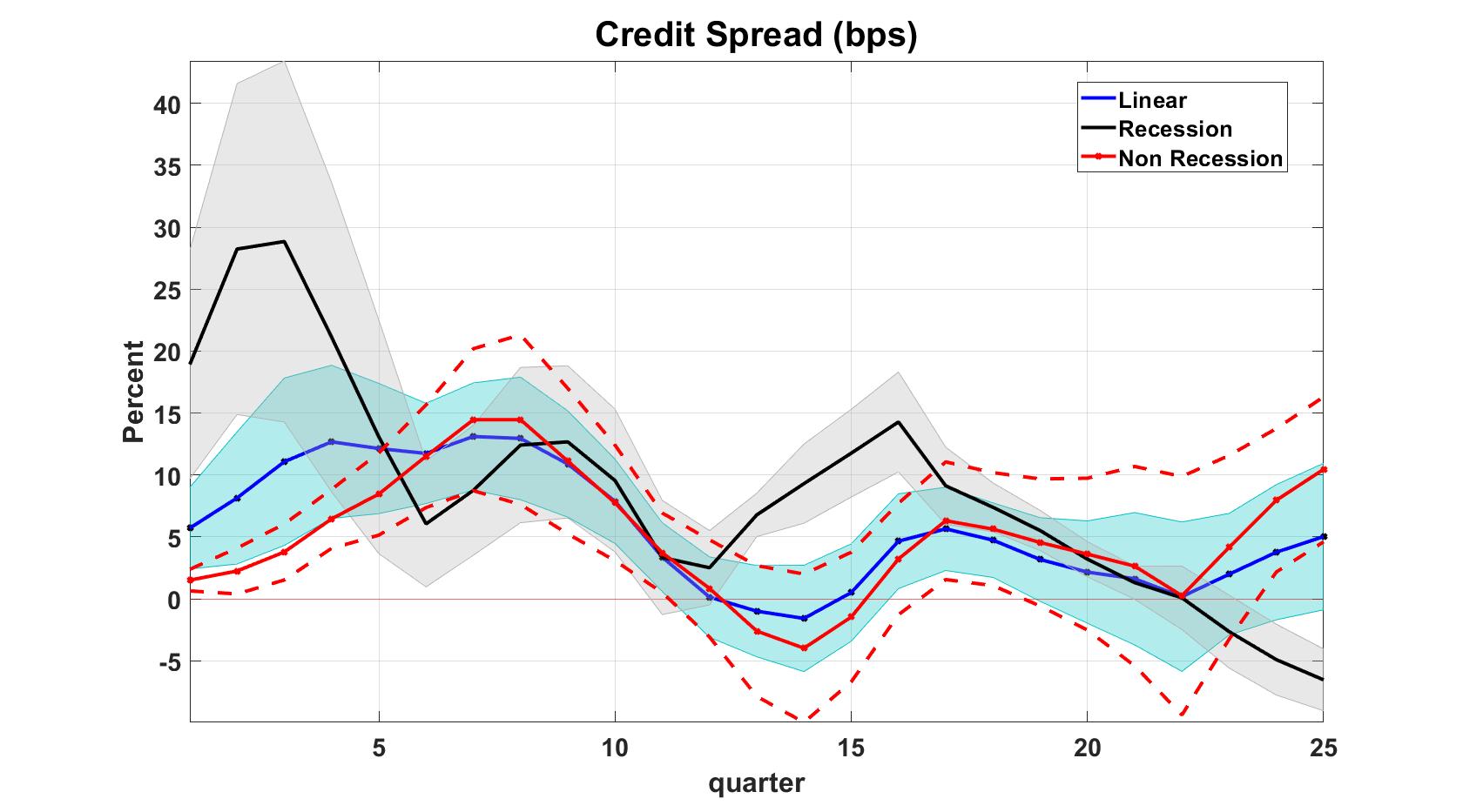} \\
			\end{array}$	}
	\end{center}
	\caption{Impact of changes in \texorpdfstring{${\eta_t^*}$}{Lg} on credit markets. Impulse responses calculated for household leverage (household credit/GDP), and the credit spread computed as the difference between the Baa and Aaa bonds. The blue line is the effect of a one standard deviation shock to \texorpdfstring{${\eta_t^*}$}{Lg} in the linear model (shaded blue area -- 68\% CI). The black line is the effect of a one standard deviation shock to \texorpdfstring{${\eta_t^*}$}{Lg} in recessions (shaded gray area -- 68\% CI). The red line is the effect of a one standard deviation shock to \texorpdfstring{${\eta_t^*}$}{Lg} in expansions (dashed red line -- 68\% CI). }
	\label{fig:CreditInd}
\end{figure}

The solid blue lines in Figures~\ref{fig:MacroVarInd} and~\ref{fig:CreditInd} show that an increase in uncertainty about credit access (after removing the effects of a shock to the first moment) has recessionary effects. The effects are amplified once state-dependence is allowed.
Once the effects across recessions (solid black line in Figures~\ref{fig:MacroVarInd} and~\ref{fig:CreditInd}) and expansions (solid red line  in Figures~\ref{fig:MacroVarInd} and~\ref{fig:CreditInd}) are decomposed, we find that the real effects of uncertainty about credit access are mainly observed in downturns. 
The effect of changes in uncertainty about credit access in expansions is negligible across the set of variables considered in the analysis. 


During downturns, macroeconomic variables, such as vehicle sales and durable consumption, that are relatively more credit-dependent, show a sharper decline compared to broader measures such as aggregate consumption. Conversely, non-durable consumption shows a significant slowdown in the periods following a shock. Investment, like the credit-sensitive components of consumption, also records a sharper slowdown. The unemployment rate peaks ten quarters after the initial shock in response. 

Qualitatively, the results from the linear model align with the empirical regularities characterizing the effects of aggregate uncertainty using alternative measures such as the VIX \citep{Bloom1}, the JLN index \citep{JLN}. The bigger effect in recessions is similar to those in \citet{Chatterjee} and reinforce the importance of state-dependence in examining the effects of uncertainty shocks.

To understand the effects on household debt and corporate borrowing costs, we examine the impulse responses of household leverage and the 3-month credit spread between Baa and Aaa corporate bonds. The conclusions are similar to those for the macroeconomic variables -- the effects are bigger during downturns. The extent of asymmetry in response in recessions is particularly prominent for the credit spread -- the peak response is a 30 basis point increase in recessions occurring within 5 quarters of the shock as opposed to a peak response of about 15 basis points in expansions occurring 8 quarters after the shock.

\subsection{Impact of an uncertainty shock with level effects \texorpdfstring{$\mathbf{\eta_t}$}{Lg}} So far, we examined impulse responses to changes in uncertainty after removing the effects of shocks to the first moment. This analysis is important as it emphasizes the importance of `pure' shocks to the second-moment characterizing access to credit. We now estimate the univariate stochastic volatility model allowing for correlation between a shock to the first and second moments.  Figure~\ref{fig:EstShocks1}  shows that  shocks to uncertainty spike during recessions when there is also a sharp decline in the growth rate of credit . The figure also shows that in expansions, uncertainty about credit access can co-move positively with credit growth (especially in the mid 1980s); however, in downturns the negative correlation is strong and consistent. 

\begin{figure}[H]
	\begin{center}	\scalebox{0.135}{
			$\begin{array}{cc}
			\includegraphics{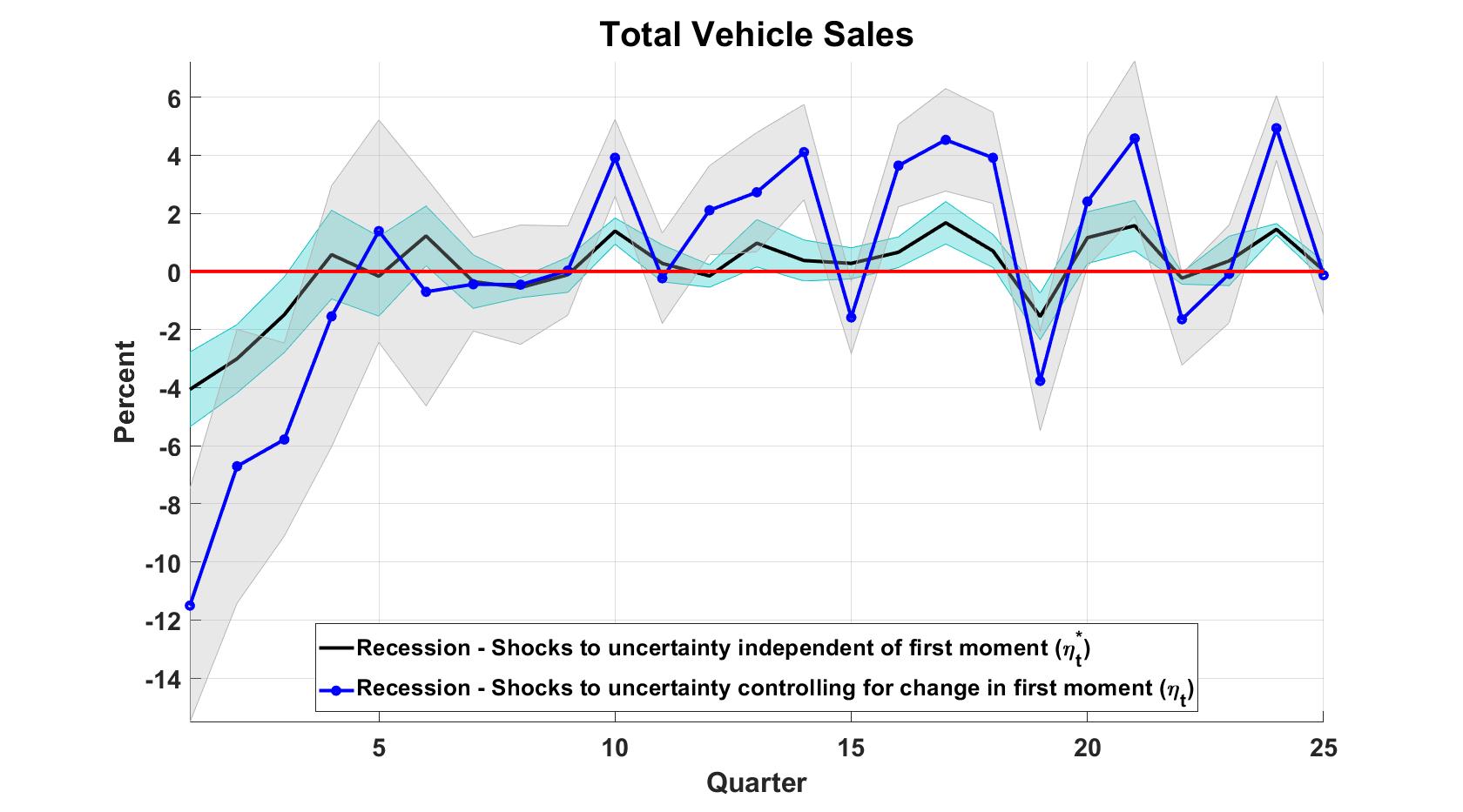} &
			\includegraphics{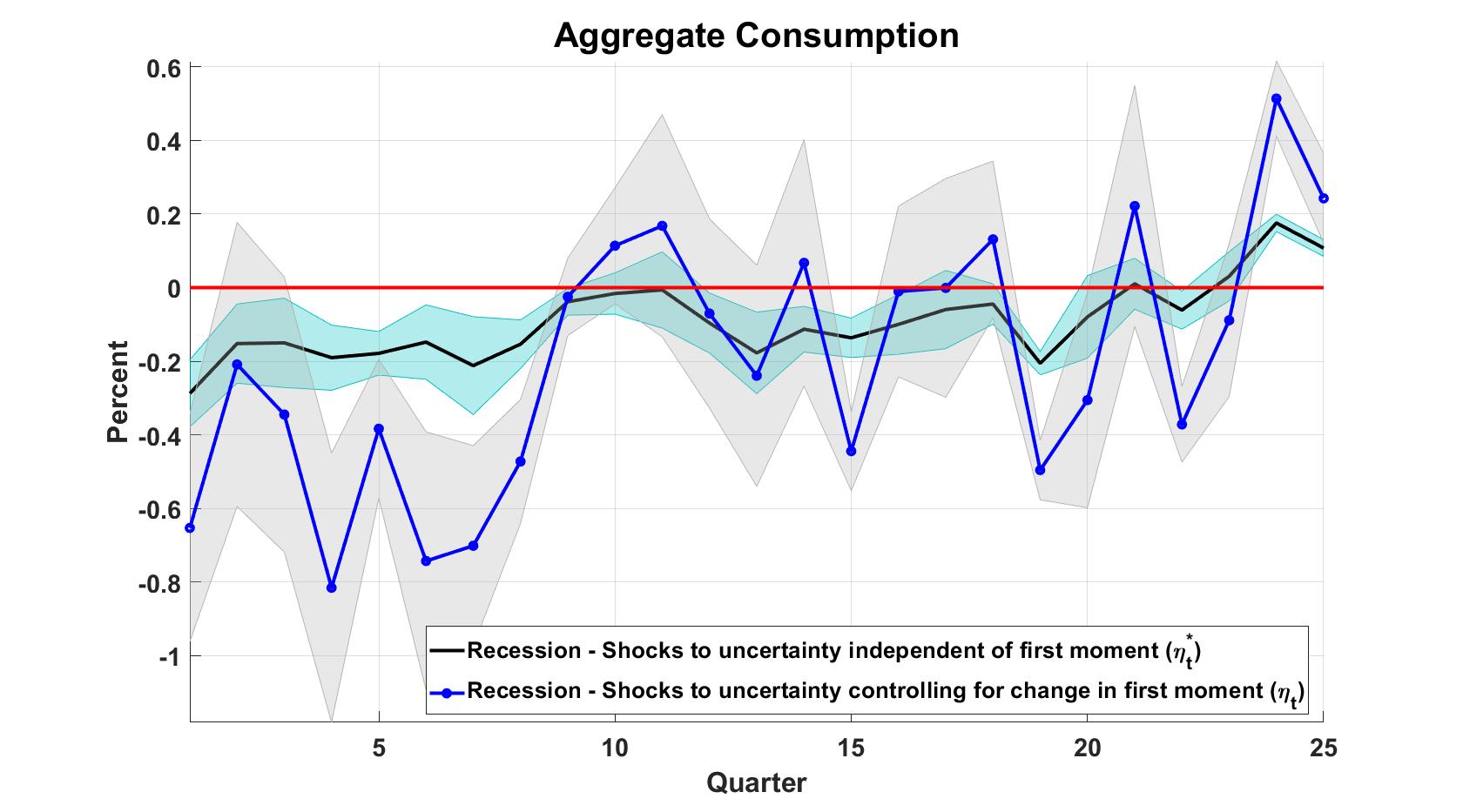}\\ 		
			\includegraphics{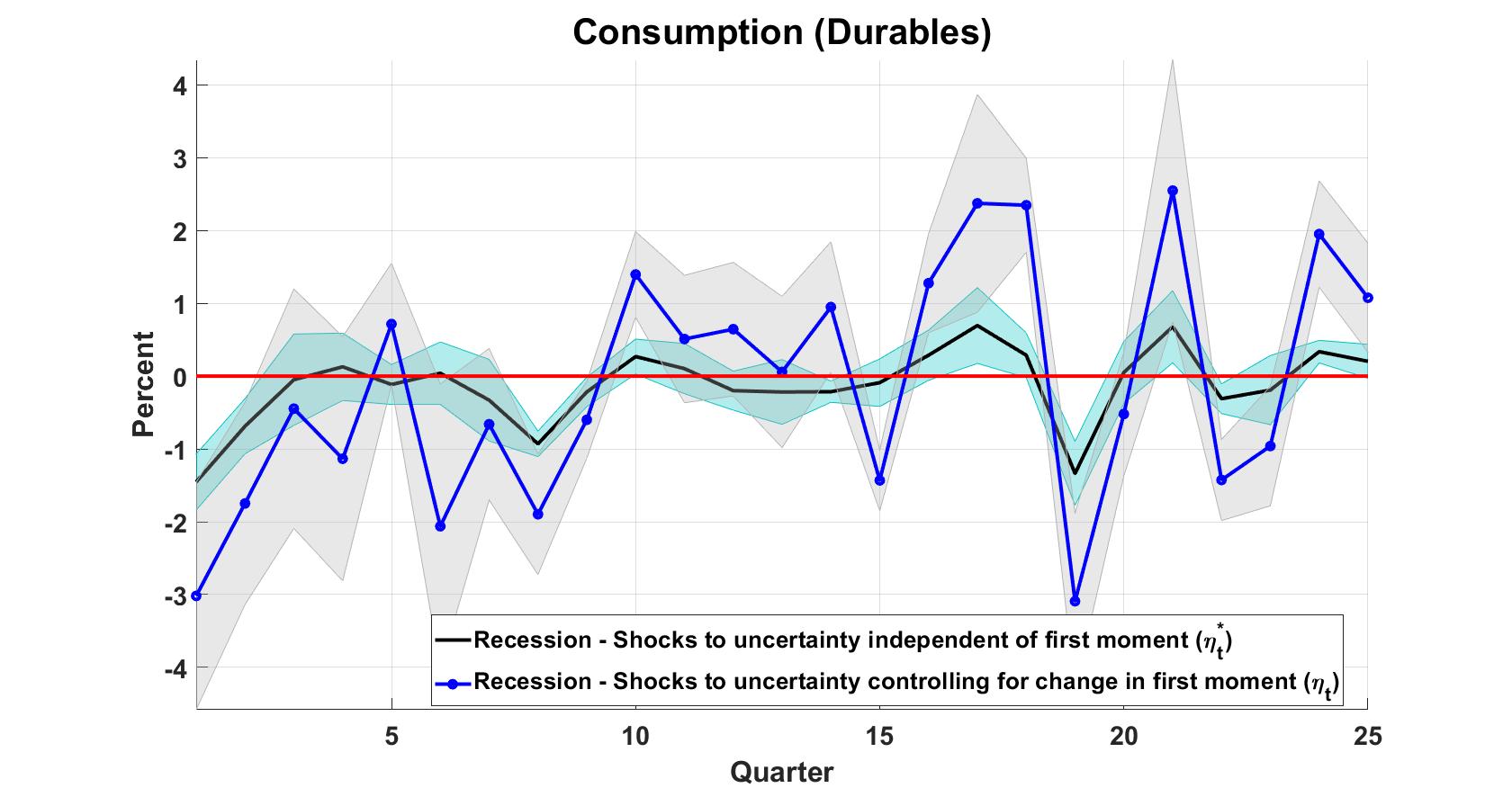} &
			\includegraphics{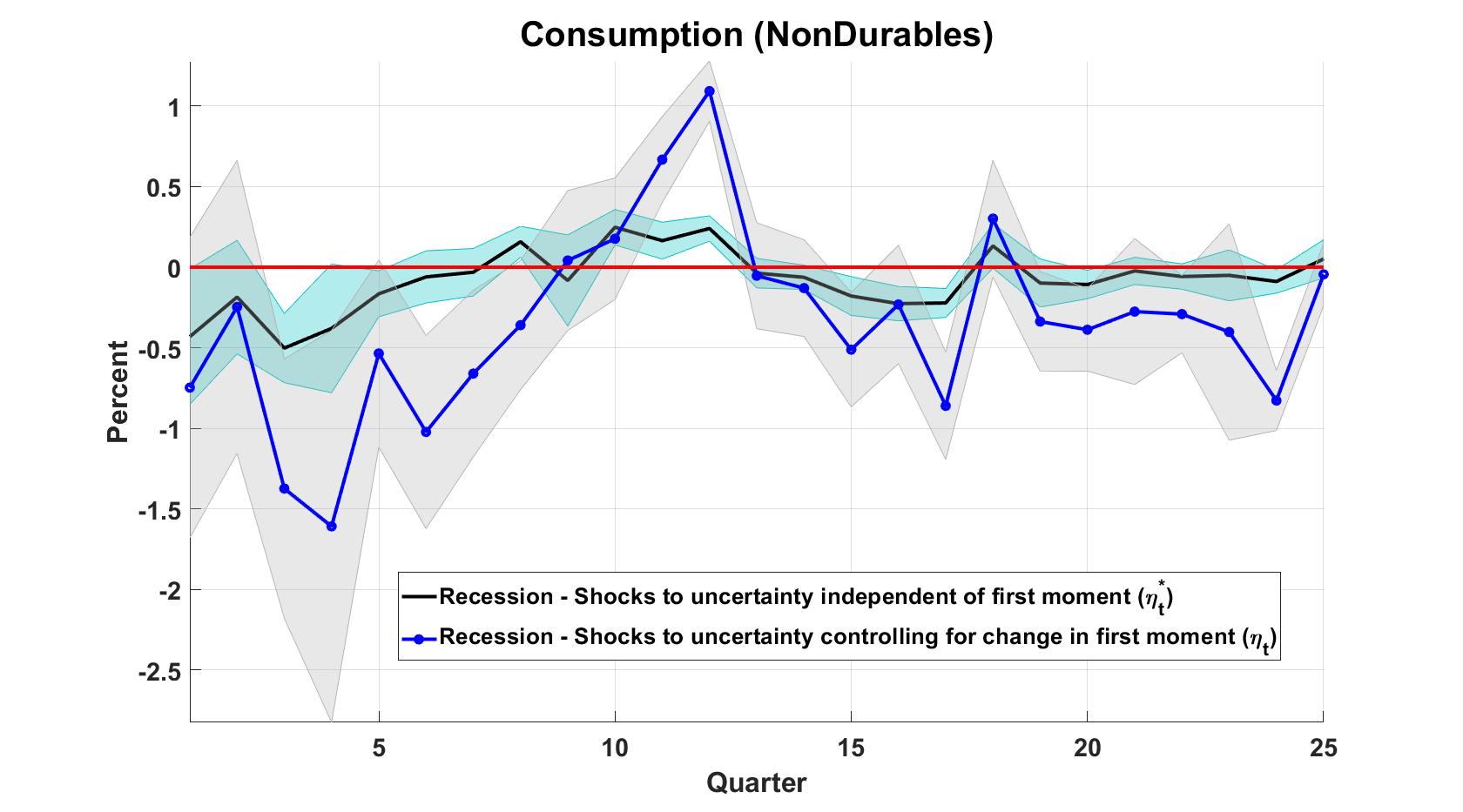}\\
			\includegraphics{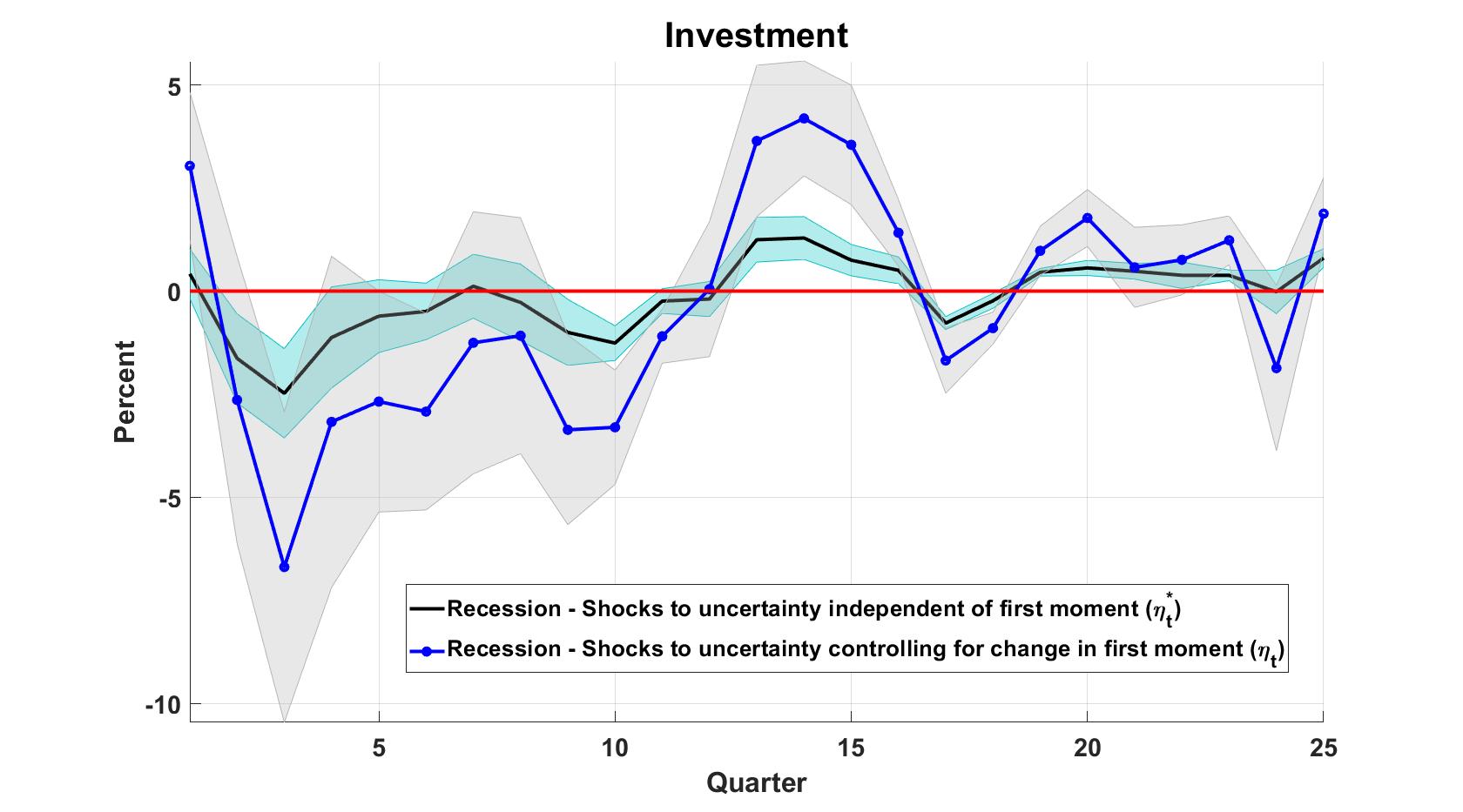} &
			\includegraphics{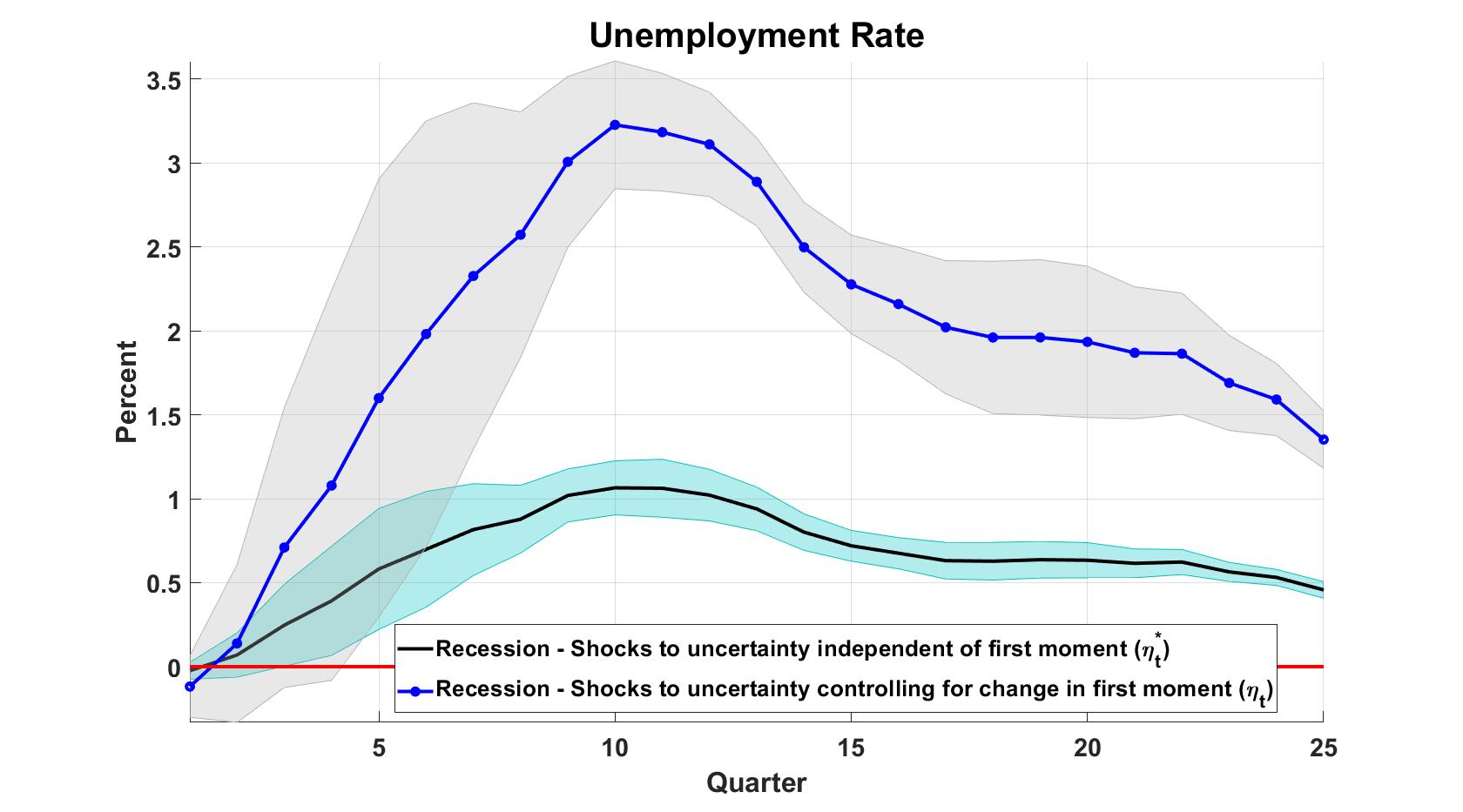}
			\end{array}$	}
	\end{center}
	\caption{Impact of changes in \texorpdfstring{${\eta_t}$}{Lg} on real activity. Impulse responses calculated for the quarterly real growth rate of total vehicle sales aggregate consumption, expenditure on durable consumption, expenditure on non-durable consumption, investment and the unemployment rate. The blue line is the effect of a one standard deviation shock to \texorpdfstring{${\eta_t}$}{Lg} in recessions (shaded gray area -- 68\% CI). The black line is the effect of a one standard deviation shock to  \texorpdfstring{${\eta_t^*}$}{Lg} in recessions (shaded blue area -- 68\% CI).}
	\label{fig:MacroVarCorr}
\end{figure}

\begin{figure}[H]
	\begin{center}	\scalebox{0.125}{
			$\begin{array}{cc}
			\includegraphics{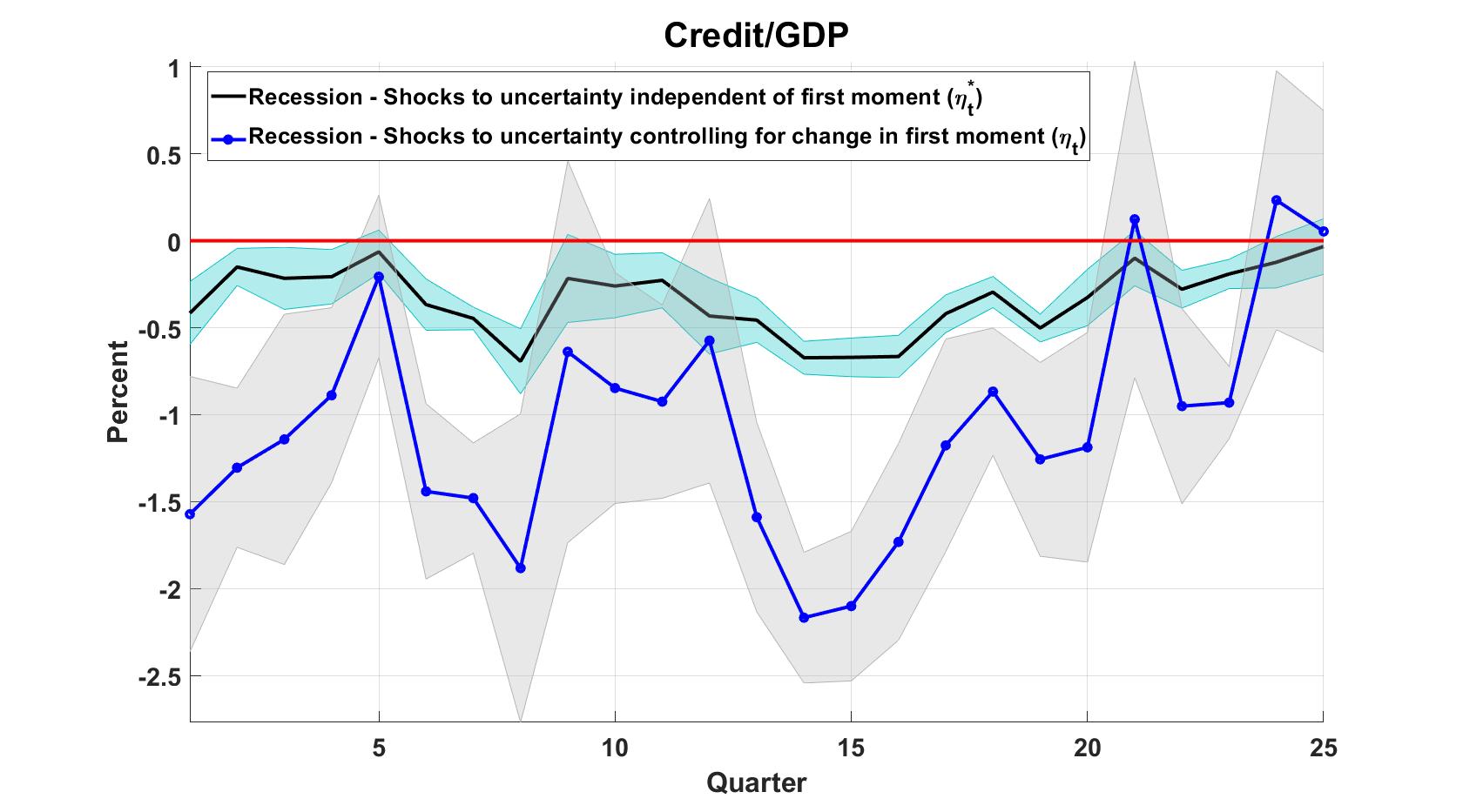} &
			\includegraphics{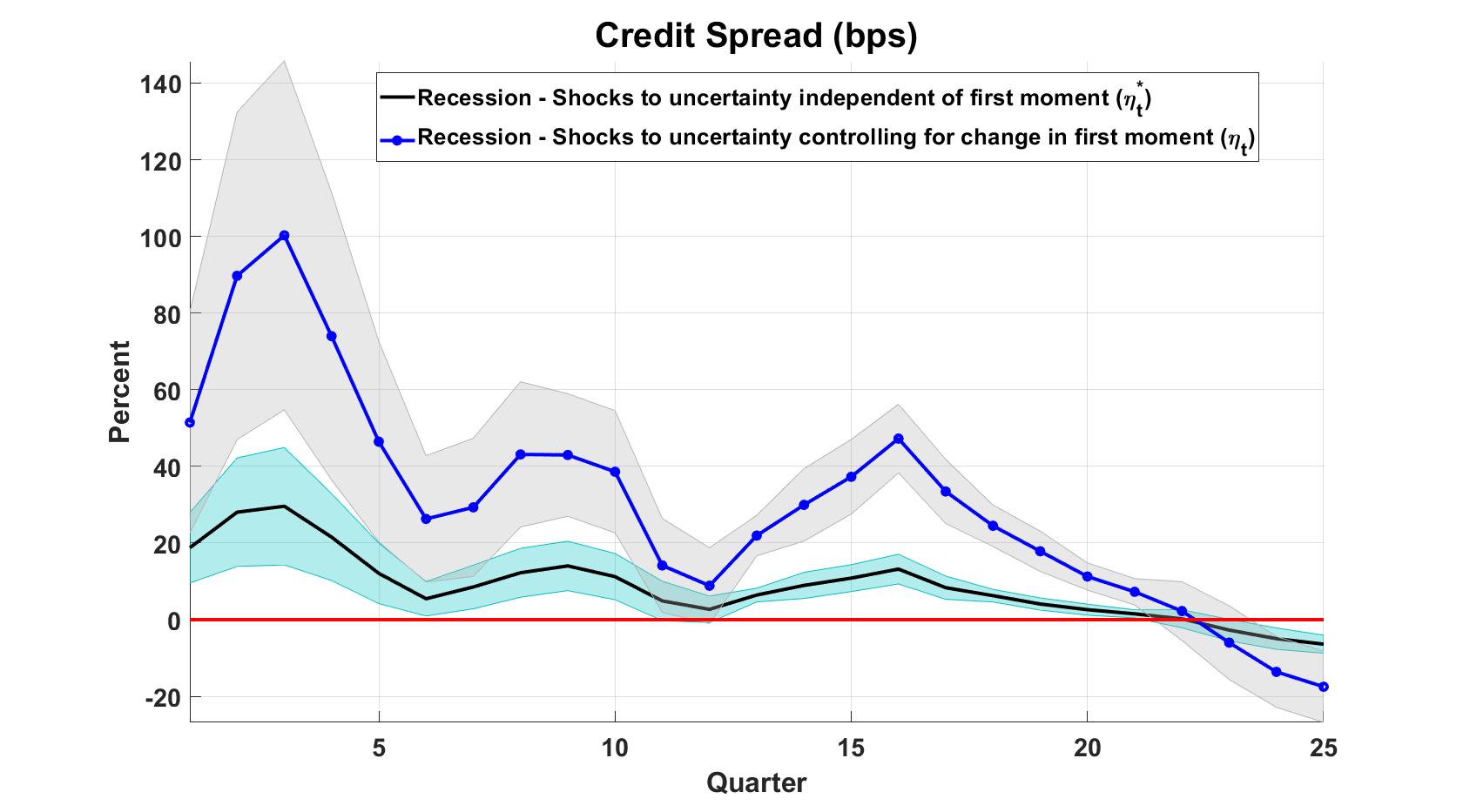} 
			\end{array}$	}
	\end{center}
	\caption{Impact of changes in \texorpdfstring{${\eta_t}$}{Lg} on credit markets. Impulse responses calculated for household leverage (household credit/GDP), and the credit spread computed as the difference between the Baa and Aaa bonds. The blue line is the effect of a one standard deviation shock to \texorpdfstring{${\eta_t}$}{Lg} in recessions (shaded gray area -- 68\% CI). The black line is the effect of a one standard deviation shock to \texorpdfstring{${\eta_t^*}$}{Lg} in recessions (shaded blue area -- 68\% CI).}
	\label{fig:CreditCorr}
\end{figure}

To fully understand the effects of uncertainty, it is necessary to take into account the nature of correlation  between the two shocks. Figures~\ref{fig:MacroVarCorr} and~\ref{fig:CreditCorr} examine the effects the shocks to uncertainty after accounting for the change in the first moment. We report the impulse responses for the recessionary regime only as our results  suggest that the impact in expansions is negligible. The solid black line in each figure captures the effects of shocks to uncertainty after removing the effects of the first moment and is identical to the results reported for the recessionary regime in Figures~\ref{fig:MacroVarInd} and ~\ref{fig:CreditInd}. The solid blue line in each figure captures the effects after allowing for interaction between the first and second moment. 
These results are particularly important, as they allow us to quantify the effects of aggregate uncertainty by considering the interaction with the first moment. While existing studies examine the time-variation of uncertainty in periods of elevated financial stress, they do not quantify the real effects by taking into account the interaction between shocks to the first and second moments. 

The decline in the credit-sensitive components of consumption, after taking into account the decline in the credit growth rate and increase in uncertainty about access to credit, is about three times higher for both the growth rate of vehicle sales and durable consumption, compared to the impact of a pure shock to uncertainty. 

As before, a significant impact is observed two periods after the shock for non-durable consumption. Across all variables, there is an amplification in the slowdown once the second-moment shock is allowed to interact with shocks to the first moment.
  
\section{Theoretical Model} We use a model of collateral constraints in the spirit of \citet*{KM} as a theoretical framework for interpreting the empirical results. In a simple real business cycle (RBC) model, a role for credit frictions is introduced by incorporating a working capital requirement on the firms' side. We assume that at the beginning of each period, firms must pay labor prior to production of output. Firms finance this labor payment with an intratemporal loan. There is no interest rate on financing the working capital because the firms pay off the loans within the period. 

\subsection{Model Description}This is a model in discrete time where agents live infinitely.  Households consume ($C_t$), supply labor ($N_t$) and save ($B_t$); $R_t$ is the gross rate of return on savings. Household utility is of CRRA type, additively separable in labor and consumption; $\gamma_c$ is inverse of the intertemporal elasticity of substitution, $\chi$ is the inverse of the Frisch elasticity, $\beta$ is the discount factor, and  $\theta$ is set such that in steady state, the hours supplied is one-third.  Households optimize:
\begin{equation*}
\underset{\{C_t,N_t,B_{t+1}\}}{\text{ max }}E_{0}\sum_{t=0}^{\infty}\beta^t\bigg(\frac{C_t^{1-\gamma_c}}{1-\gamma_c}- \theta\frac{N_t^{1+\chi}}{1+\chi}\bigg)
\end{equation*}
subject to
\begin{equation*}
C_t+B_{t+1}=W_t{N_t}+R_{t-1}B_{t}+\Pi_t\text{ ;}
\end{equation*}
$W_t$ is the real wage and $\Pi_t$ denotes residual profits from firms. The first order conditions of the household are 
\begin{equation}
{C_t^{-\gamma_c}} =\lambda_t\text{ ;}
\end{equation} 
\begin{equation}
\theta{N_t^{\chi}} ={\lambda_t}W_t\text{ ;}
\end{equation} 
\begin{equation}
1={\beta}E_t\Big[\Big(\frac{C_{t+1}}{C_t}\Big)^{-\gamma_c}{R_t}\Big]\text{ .}
\end{equation}
The firms own capital stock ($K_t$) and hire labor input ($N_t$) from households to produce output $Y_t$ using the Cobb-Douglas production technology 
\begin{equation}
Y_{t}=A_tK_t^\alpha{N_t}^{1-\alpha}\text{ ;}
\end{equation}
$A_t$ describes the state of aggregate technology. Capital accumulation is subject to adjustment costs and evolves as
\begin{equation}
K_{t+1}=(1-\delta)K_t+I_t-\frac{\phi}{2}{\Big(\frac{I_t}{K_t}-\delta\Big)^2}K_t \text{ ;}
\end{equation}
$I_t$ is investment, $\delta$ is the rate at which capital depreciates and $\phi$ is the cost of adjusting changes in capital. In this framework, investment is financed using residual profits (dividends). For now, we assume that firms do not issue intertemporal debt. The dividend payout is
\begin{equation*}
\Pi_t=Y_t-w_tN_t-I_t\text{ .}
\end{equation*}
 We introduce a working capital constraint in this simple RBC set-up and  assume that firms must pay labor prior to producing output and  the working capital is financed through an intratemporal loan. The amount the firm can borrow is limited to a fraction $\zeta_t$ of the total value of capital, with $q_t$ the price of capital. We abstract from intertemporal debt in the baseline model. The intuition governing the impact of an uncertainty shock remains unchanged compared to the scenario with intertemporal debt.
 The constraint restricting working-capital financing is
\begin{equation*}
W_tN_t \leq \zeta_tq_tK_t
\end{equation*}
 The optimization problem of the firm, given the working capital constraint is
\begin{equation*}
\underset{\{I_t,N_t,K_{t+1}\}}{\text{max }} {E_{0}\sum_{t=0}^{\infty}\beta^t\frac{\lambda_{t}}{\lambda_0}}\Big[A_tK_t^\alpha{N_t}^{1-\alpha}-W_tN_t-I_t\Big] \text{ ,}
\end{equation*}
subject to
\begin{equation*}
K_{t+1}=(1-\delta)K_t+I_t-\frac{\phi}{2}{(\frac{I_t}{K_t}-\delta)^2}K_t \text{ ,}
\end{equation*} 
and
\begin{equation*}
W_tN_t \leq \zeta_tq_tK_t \text{ .}
\end{equation*}
Firm dividends at time $t+j$ are discounted using the stochastic discount factor from the optimization problem of the households -- $\beta^j{\lambda_{t+j}}/{\lambda_t}$. Let $\mu_t$ be the multiplier on the working capital constraint; $\mu_t >0$ implies that the constraint binds and changes in $\zeta_t$  impact the equilibrium conditions of the model. The first order conditions corresponding to the optimal choices of investment ($I_t$), labor ($N_t$) and capital ($K_{t+1}$) are  
\begin{equation}
W_t(1+\mu_t)=(1-\alpha)A_tK_t^{\alpha}N_t^{-\alpha} \text{ ,}
\end{equation}
\begin{equation}
1=q_t[1-\phi(\frac{I_t}{K_t}-\delta)] \text{ ,}
\end{equation}
\begin{equation}
q_t={\beta}E_t\frac{\lambda_{t+1}}{\lambda_t}\Bigg[{\alpha}A_{t+1}K_{t+1}^{\alpha-1}N_{t+1}^{1-\alpha} +q_{t+1} \Bigg( (1-\delta)  + \mu_{t+1}\zeta_{t+1} -\frac{\phi}{2}\Big(\frac{I_{t+1}}{K_{t+1}}-\delta\Big)^2 +\phi\Big(\frac{I_{t+1}}{K_{t+1}}-\delta\Big)\Bigg)     \Bigg] \text { ,}
\end{equation}
and\begin{equation}
\mu_t(W_tN_t - \zeta_tq_tK_t)=0 \text{ with } \mu_t \geq 0 \text { .}
\end{equation}
When the credit constraint binds, $\mu_t >0$ and $(W_tN_t - \zeta_tq_tK_t)=0$; when the constraint is slack,  $\mu_t=0$ and $(W_tN_t - \zeta_tq_tK_t)<0$.
The market clearing condition is 
\begin{equation}
Y_t=C_t+I_t \text { .}
\end{equation}

We assume that the fraction of the value of the firm that can be borrowed follows an AR(1) process with a time-varying second moment,
\begin{equation}
\begin{aligned}
& \zeta_t-\zeta_{ss}={\rho}(\zeta_{t-1}-{\zeta_{ss}})+ \exp{(h_t/2)}\epsilon_{t} \text { ,}\\
& h_t=(1-\rho_h)\overline{h}+\rho_h{h_{t-1}}+\tau{\eta_t^*} \text { ;}
\end{aligned}
\end{equation}
$\zeta_{ss}$ is the fraction the firm can borrow against the value of capital in steady state; $\overline{h}$ is the average uncertainty in credit availability; and $\tau$ is the extent of stochastic volatility in credit availability. 

This simple set-up captures the essence of the transmission of shocks to volatility about credit availability. A shock to the first moment $\epsilon_{t}$ is an exogenous increase in credit availability, whereas a shock to ${\eta_t^*}$ acts like a mean-preserving spread that implies an increase in uncertainty about credit availability. The specification of the stochastic volatility process is exactly the same as in the model presented in section 2. The empirical section  uses the growth rate of credit available to the nonfinancial sector to estimate Equation~\eqref{observationequationbasic}. 
While examining shocks to uncertainty, we set the correlation between the first and second moment shock to zero to allow independent effect of changes in uncertainty. Equations (13)-(21) summarize the equilibrium conditions in the model.

\subsection{Model Calibration} The model is calibrated quarterly and the behavioral parameters of the model are standard. The share of capital in the Cobb-Douglas production function is set at $\alpha=0.33$, the discount factor $\beta=0.99$, the rate of depreciation of capital $\delta=0.02$, and investment adjustment costs $\phi=4$. The parameter $\theta$ scaling the disutility of labor supply is set to 5.7241 so that labor hours in steady state are $\frac{1}{3}$. The inverse interemporal elasticity of substitution is fixed at $\gamma_c=1$ and the inverse of the Frisch elasticity of labor supply at $\chi=1$.

To fix the parameters governing the stochastic volatility characterizing access to credit, we revisit the posterior estimates presented in Table \ref{tab:Posterior-Mean-Estimates SV} of section 3. We set $\overline{h}=\overline{\mu_h}=-10.12$, $\rho_h=\phi_h=0.91$, $\tau=0.25$ and $\rho=\phi_{y}=0.83$, corresponding to the posterior means obtained from the estimates in section 3. 

The steady state value of $\zeta$ limiting the firm's borrowing  is now discussed. When the constraint binds in steady state, $\mu>0$. This imposes certain restrictions on the steady state values of $\zeta=\zeta_{ss}$. For $\mu >0$ in steady state, it can be shown that $\zeta_{ss}< \frac{1-\alpha}{\alpha}\Big(\frac{1}{\beta}-(1-\delta)\Big)$. Given the values of $\alpha, \beta, \delta$, we find that $\zeta_{ss} < 0.0602$. To fix the value of $\zeta_{ss}$, we consider the average growth rate of credit to the nonfinancial sector. It is 1.7\% in the sample spanning from $1978Q1-2018Q4$, so the constraint binds in steady state  and is well below the required limit.

\subsection{Model Solution}The goal of this paper is to explore the effects of a change in the second moment that captures uncertainty shocks to credit constraints. A first order approximation shuts down the effects of changes in higher-order moments by construction. A second-order solution  impacts  expected values but does not influence the dynamics as higher-order terms do not independently enter the solution \citep*{SGU}. It is necessary to consider a third-order approximation for uncertainty to have dynamic effects. We do so by using the perturbation methods combined with pruning (to prevent explosive solutions) in \citet*{AFVVRR}. The constraint always binds.

\subsection{Transmission of shock to uncertainty}  The behavior of the firm in this environment is subject to collateral constraints on borrowing. Since the constraint limits the quantity a firm may borrow, it has implications for how much labor the firm can hire. That is, it introduces a wedge between the perfectly competitive equilibrium and the realized outcome in the constrained economy. This simple assumption generates important implications by introducing a distortion through the intratemporal optimization condition of the households in the optimal labor-leisure trade-off. In the absence of sticky prices, this wedge can generate a simultaneous decline across variables in the model economy.

\citet*{BasuBundick} show that in a flexible-price model, an uncertainty shock introduced through time variation in volatility of aggregate demand, cannot generate a simultaneous decline in consumption, investment, and output; this is regarded as a stylized fact that characterizes the effect of an uncertainty shock. \citet*{BasuBundick} subsequently highlight the importance of nominal rigidities through the wedge introduced by the presence of a mark-up in labor demand. 

The collateral constraint in our model, operating through the intratemporal working capital channel, distorts the flexible price equilibrium. The labor demand side of the model now reflects the effect of the collateral constraint:\begin{equation*}
W_t(1+\mu_t)=(1-\alpha)A_tK_t^{\alpha}N_t^{-\alpha} \text{ ;}
\end{equation*}
the labor supply side is unchanged:
\begin{equation*}
W_t=\frac{\theta{N_t^\chi}}{C_t^{-\gamma_c}} \text{.}
\end{equation*}
An uncertainty shock in the model triggers a precautionary response, leading to an increase in marginal utility and an increase in labor supply. In a flexible price model, this increase in labor supply is absorbed and thus prevents the simultaneous decline in consumption, investment and output. In a model with collateral constraints, when there is an uncertainty shock to $\zeta_t$ that limits that amount that a firm can borrow, the multiplier on the borrowing constraint now provides the additional margin provided by the presence of the mark-up in a model with sticky prices. When there is an exogenous shock to uncertainty about access to credit -- here interpreted as a shock to the second moment of $\zeta_t$, there is an increase in $\mu_t$, the Lagrange multiplier on the binding constraint, which prevents labor demand from increasing and in turn generates a decline in output, consumption and investment. 

To better understand  how the precautionary motives interact with the endogenous variables, we now examine the expression for the third order accurate solution of the model. Following \citet*{AFVVRR}, the third order solution after pruning is
\small{
\begin{multline*}
x^{3rd}_t=h_v\begin{bmatrix} 
x^{3rd}_{t-1} \\  \mathbf{0} 
\end{bmatrix} +
2H_{vv}\Bigg(\begin{bmatrix} 
x^{f}_{t-1} \\  \epsilon^\zeta_t \\ \eta_t^* 
\end{bmatrix} \otimes \begin{bmatrix} 
x^{f}_{t-1} \\  \epsilon^\zeta_t \\ \eta_t^* 
\end{bmatrix}\Bigg)+ \\
H_{vvv}\Bigg(\begin{bmatrix} 
x^{f}_{t-1} \\  \epsilon^\zeta_t \\ \eta_t^* 
\end{bmatrix} \otimes \begin{bmatrix} 
x^{f}_{t-1} \\  \epsilon^\zeta_t \\ \eta_t^* 
\end{bmatrix}\otimes \begin{bmatrix} 
x^{f}_{t-1} \\  \epsilon^\zeta_t \\ \eta_t^* 
\end{bmatrix}\Bigg) +
\frac{3}{6}h_{\sigma\sigma{v}}\begin{bmatrix} 
x^{f}_{t-1} \\  \epsilon^\zeta_t \\ \eta_t^*  
\end{bmatrix} +\frac{1}{6}h_{\sigma\sigma\sigma}\sigma^2
\end{multline*}}
The vector of state variables $x_{t}$ is  $[K_t, \sigma_t, \zeta_t]$ for our model,  and the vector of shocks is $\epsilon_{t}=[\epsilon_{t}^\zeta,\eta_t^*]$. The extended state space vector for higher-order solution, $v_t$, includes the second and third order terms along with the usual first-order effects.\footnote{The extended state space vector $v_t$, is given as follows: $\Bigg[
	x^{f}_{t-1},\epsilon^\zeta_t,\eta_t^*,
	x^{f}_{t-1} \otimes x^{f}_{t-1},  x^{f}_{t-1} \otimes \epsilon^\zeta_t,  x^{f}_{t-1} \otimes \eta_t^*, \epsilon^\zeta_t\otimes\epsilon^\zeta_t, \epsilon^\zeta_t\otimes\eta_t^*, \eta_t^*\otimes\eta_t^*,
	x^{f}_{t-1} \otimes x^{f}_{t-1}\otimes x^{f}_{t-1},
	 x^{f}_{t-1} \otimes x^{f}_{t-1}\otimes \epsilon^\zeta_t, 
	 x^{f}_{t-1} \otimes x^{f}_{t-1}\otimes \eta_t^*,   
	 x^{f}_{t-1} \otimes \epsilon^\zeta_t\otimes \epsilon^\zeta_t, 
	 x^{f}_{t-1} \otimes \epsilon^\zeta_t\otimes \eta_t^*,
	  x^{f}_{t-1} \otimes \eta_t^*\otimes \eta_t^*, 
	 \epsilon^\zeta_t\otimes \epsilon^\zeta_t\otimes \epsilon^\zeta_t, \epsilon^\zeta_t\otimes \epsilon^\zeta_t\otimes \eta_t^*, \epsilon^\zeta_t\otimes \eta_t^*\otimes \eta_t^*,
	  \eta_t^*\otimes \eta_t^*\otimes \eta_t^*
\Bigg]$.$\otimes$ is the Kronecker product.} The matrices $H_{vv}$ and $H_{vvv}$ summarize  the coefficients corresponding to the interaction terms at the second and third order, respectively; $h_{\sigma\sigma\nu}$ captures the direct effect of an uncertainty shock for a third-order approximation; and $h_{\sigma\sigma\sigma}$ is the adjustment to the non-stochastic steady state at the third-order.

A change in $\eta_t^*$ captures an uncertainty shock in the model. Using the third order solution, we decompose the impulse response on impact to an uncertainty shock about access to credit into two parts; a direct effect and an interaction effect. 
On impact, an uncertainty shock has a direct effect on $\sigma_t$ through $\eta_t^*$, and the size of the impact is obtained by the relevant coefficient of the matrix $h_{\sigma\sigma{v}}$.\footnote{In our model $\dim(h_{\sigma\sigma{v}})=5\times5$. So the fifth column of this matrix captures the non-zero effects of an uncertainty shock.} Column 1 of Table~\ref{tab:Decompose} reports direct effects.

We now isolate the interaction effects. Under a rational expectations assumption, agents can observe the shock to uncertainty about credit-access $\eta_t^*$ at time $t$, hence, $E_t\eta_t^*=\eta_t^*$. If the impulse responses are computed at the unconditional mean, then on impact all $(t-1)$ dated variables are 0, that is, $x^{f}_{t-1}=0$. Additionally, the interaction terms of the form $E_t(x^{f}_{t-1}\epsilon^\zeta_t\eta_t^* )$, $E_t(x^{f}_{t-1}\epsilon^\zeta_t\epsilon^\zeta_t)$, $E_t(x^{f}_{t-1}\eta_t^* \eta_t^* )$, and $E_t(x^{f}_{t-1}\eta_t^*\epsilon^\zeta_t)$ can be expressed as $\eta_t^*E_t(x^{f}_{t-1}\epsilon^\zeta_t )$, $E_t(x^{f}_{t-1}\epsilon^\zeta_t\epsilon^\zeta_t)$, ${\eta_t^*}^2E_t(x^{f}_{t-1} )$, and $\eta_t^*E_t(x^{f}_{t-1}\epsilon^\zeta_t)$;
these terms are zero since the shocks to the first moment are uncorrelated with the $x^{f}_{t-1}$; the remaining interaction terms are  $E_t(\epsilon^\zeta_t\epsilon^\zeta_t\epsilon^\zeta_t )$, $E_t(\epsilon^\zeta_t\epsilon^\zeta_t\eta_t^*)$, $E_t(\epsilon^\zeta_t\eta_t^*\eta_t^*)$ and $E_t(\eta_t^*\eta_t^*\eta_t^*)$. 

A shock to $\eta_t^*$ implies that the effects on the interaction terms are $E_t(\epsilon^\zeta_t\epsilon^\zeta_t\epsilon^\zeta_t )$, $\eta_t^*E_t(\epsilon^\zeta_t\epsilon^\zeta_t)$,  $(\eta_t^*)^2E_t(\epsilon^\zeta_t)$ and $(\eta_t^*)^3$.  Given that $\epsilon_t^\zeta \text{ } \underset{iid}{\sim}\text{ }   (0,1)$, $\eta_t^* \text{ } \underset{iid}{\sim}\text{ }   (0,1)$, $E_t(\eta_t^*\epsilon^\zeta_t)=0$, $E_t\epsilon^\zeta_t=0$, and $E_t(\epsilon^\zeta_t\epsilon^\zeta_t)=1$, the non-zero interaction effect in the extended state space shows up for $\eta_t^*E_t(\epsilon^\zeta_t\epsilon^\zeta_t)$. The interaction effects are therefore captured by the coefficients corresponding to the state variable $\eta_t^*E_t(\epsilon^\zeta_t\epsilon^\zeta_t)$ in the  $H_{vvv}$ matrix. Column 2 of Table~\ref{tab:Decompose} reports the effect on impact attributed to the interaction channel. The effect on the control variables is similarly  obtained, from,
\small{
\begin{multline*}
y^{3rd}_t=g_v\begin{bmatrix} 
x^{3rd}_{t-1} \\  \mathbf{0} 
\end{bmatrix} +
2G_{vv}\Bigg(\begin{bmatrix} 
x^{f}_{t-1} \\  \epsilon^\zeta_t \\ \eta_t^* 
\end{bmatrix} \otimes \begin{bmatrix} 
x^{f}_{t-1} \\  \epsilon^\zeta_t \\ \eta_t* 
\end{bmatrix}\Bigg)+ \\
G_{vvv}\Bigg(\begin{bmatrix} 
x^{f}_{t-1} \\  \epsilon^\zeta_t \\ \eta_t^* 
\end{bmatrix} \otimes \begin{bmatrix} 
x^{f}_{t-1} \\  \epsilon^\zeta_t \\ \eta_t^* 
\end{bmatrix}\otimes \begin{bmatrix} 
x^{f}_{t-1} \\  \epsilon^\zeta_t \\ \eta_t^* 
\end{bmatrix}\Bigg) +
\frac{3}{6}h_{\sigma\sigma{v}}\begin{bmatrix} 
x^{f}_{t-1} \\  \epsilon^\zeta_t \\ \eta_t^*  
\end{bmatrix} +\frac{1}{6}g_{\sigma\sigma\sigma}\sigma^2
\end{multline*}}
Column 3 of Table~\ref{tab:Decompose} summarizes the total effect $=$ the sum of the direct effect and interaction effect. An advantage of this decomposition is that it helps isolate the precautionary-driven change in hours supplied and the contraction in hours demanded that works through the wedge $\mu_t$. 

\begin{figure}[H]
	\begin{center}	{
			$\begin{array}{c}
			\includegraphics[width=120mm,height=90mm,angle=0]{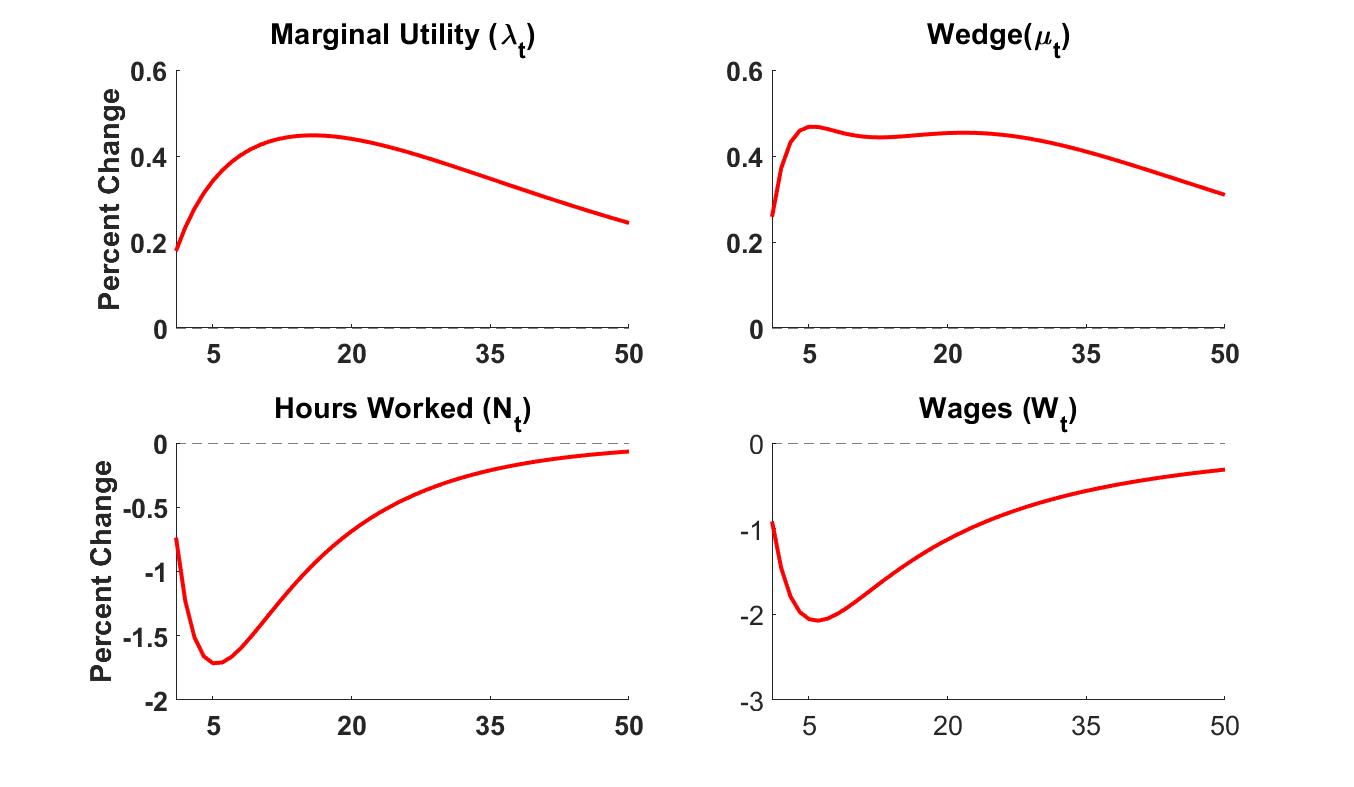} 
			\end{array}$	}
	\end{center}
	\caption{Impulse responses for a 1 standard deviation shock to uncertainty $\eta_t^{*}$  in the DSGE model. An uncertainty shock to credit access triggers an increase in the wedge between labor demand and labor supply that leads to a decline in hours and wages in equilibrium.}
	\label{fig:IRFsModel1}
\end{figure}

\begin{figure}[H]
	\begin{center}	{
			$\begin{array}{c}
			\includegraphics[width=120mm,height=60mm,angle=0]{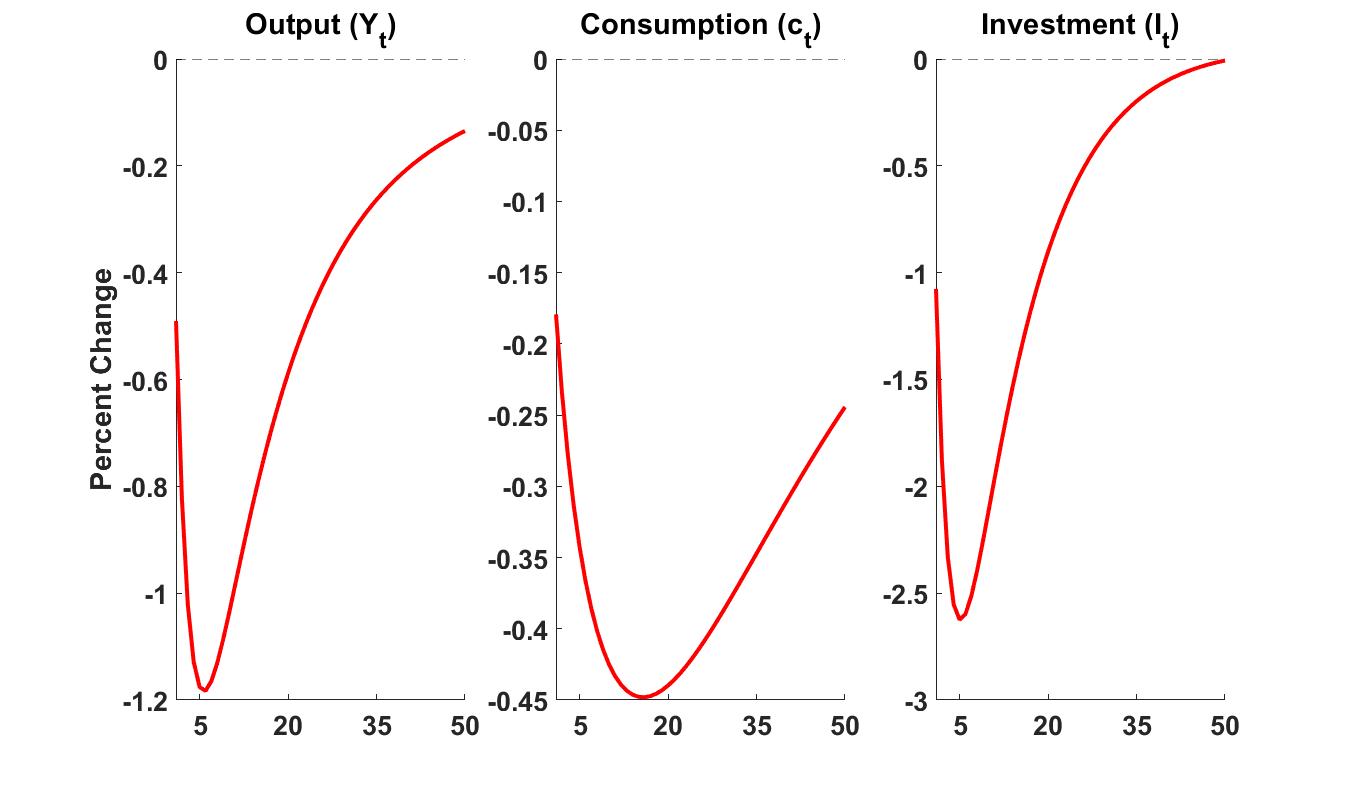} 
			\end{array}$	}
	\end{center}
	\caption{Impulse responses for a 1 standard deviation shock to uncertainty $\eta_t^{*}$  in the DSGE model. An uncertainty shock to credit access triggers a simultaneous decline in consumption, investment, output and hours.}
	\label{fig:IRFsModel2}
\end{figure}

Column 1 of Table~\ref{tab:Decompose} shows the direct effects; clearly an uncertainty shock generates a precautionary driven decline in consumption, and an increase in the marginal utility and labor supply. However, there is also an increase in the wedge through the direct effect. Column 2 of Table~\ref{tab:Decompose} shows that the interaction effects between these different elements, dampens the total hours supplied, GDP and investment, but amplifies the initial decline in consumption from the precautionary response. Column 3 of Table~\ref{tab:Decompose} presents total effects, and correspond to the response on impact in Figures~\ref{fig:IRFsModel1} and ~\ref{fig:IRFsModel2}.
\begin{table}[H]
	\caption{Decomposing the effects of an uncertainty ($\eta_t^{*}$) shock on impact}
	\label{tab:Decompose}\small{
	\begin{center}
	\begin{tabular}{p{0.28\textwidth} p{0.225\textwidth} p{0.225\textwidth}p{0.12\textwidth} }\hline\hline
	Variables &	Direct Effect from $\eta_t^*$ (\% change)& Interaction Effect from $\eta_t^*E_t(\epsilon_t^\zeta,\epsilon_t^\zeta)$ (\% change)&Total Effect (\% change)\\
	 & (1) &   (2)  &  (3)\\	
		 \hline 
Consumption                                                                             & -0.099                                                                             & -0.081                                                                          & -0.179                                                                    \\
Real Wages                                                                              & -0.040                                                                             & -0.875                                                                          & -0.915                                                                    \\
Rate of Return on Capital                                                               & -0.087                                                                             & 0.009                                                                           & -0.079                                                                    \\
GDP                                                                                     & 0.039                                                                              & -0.529                                                                          & -0.491                                                                    \\
Hours & 0.058                                                                              & -0.794                                                                          & -0.736                                                                    \\
Marginal Utility                                                                        & 0.099                                                                              & 0.081                                                                           & 0.179                                                                     \\
Wedge                                                                                   & 0.046                                                                              & 0.213                                                                           & 0.258                                                                     \\
Price of Capital                                                                        & 0.018                                                                              & -0.063                                                                          & -0.045                                                                    \\
Investment                                                                              & 0.226                                                                              & -1.302                                                                          & -1.076                                                               \\  \hline\hline
\end{tabular}
\end{center}}
\end{table}
These results operate through the firm's binding borrowing constraint and are insightful as an uncertainty shock of this form generates a simultaneous decline in consumption, investment, and output in a flexible price environment. The model predictions thus match the qualitative patterns recovered from empirically analyzing the effects of shocks to credit access, as well as building intuition about the cause of the slowdown in activity. 

 \subsection{Extension with rule-of-thumb households}

The impulse responses in section 5.4 show that an uncertainty shock about access to credit is recessionary.  In the baseline specification, households can engage in precautionary savings and this is an important element in the transmission of an uncertainty shock. We consider a simple extension of the framework where  heterogeneity is introduced in the description of households. Within the continuum of infinitely-lived households we introduce a fraction $\nu$ that do not have access to savings, and therefore do not own any assets or liabilities and in each period consume their labor income. These are the rule-of-thumb consumers or the non-Ricardian households. The remaining $(1-\nu)$ fraction of households are able to save as before. The non-Ricardian households are important as they cannot engage in precautionary behaviour. We examine the consumption response of the non-Ricardian households to understand if they  are more vulnerable to the effects of uncertainty in credit access in the macroeconomy. 

The equilibrium conditions of the model are now updated to reflect this household heterogeneity. The behavior of the Ricardian households is identical to those in the baseline model in Section 5.1. The rule-of-thumb/non-Ricardian agents maximize period utility 
\begin{equation*}
\underset{\{C_t,N_t\}}{\text{ max }}\bigg(\frac{{C_t^{NR}}^{1-\gamma_c}}{1-\gamma_c}- \theta_R\frac{{N_t^{NR}}^{1+\chi}}{1+\chi}\bigg)
\end{equation*} subject to  \begin{equation*}C_t^{NR}=W_tN_t^{NR}\text{ .}\end{equation*}
The first order conditions give
\begin{equation}
N_t^{NR}=\Bigg(\frac{W_t^{1-\gamma_c}}{\theta_R}\Bigg)^\frac{1}{\chi+\gamma_c}\text{ .}
\end{equation}
The parameter $\theta_R$ is chosen such that hours supplied by the non-Ricardian agents in steady state is ${1}/{3}$. Given, $\gamma_c=1$, Equation ($24$) implies that hours supplied by the non-Ricardian households is fixed. The implications of this feature are seen when impulse responses of hours are examined for an uncertainty shock to $\zeta$.

\begin{figure}[H]
	\begin{center}	{
			$\begin{array}{c}\scalebox{0.225}{
			\includegraphics[]{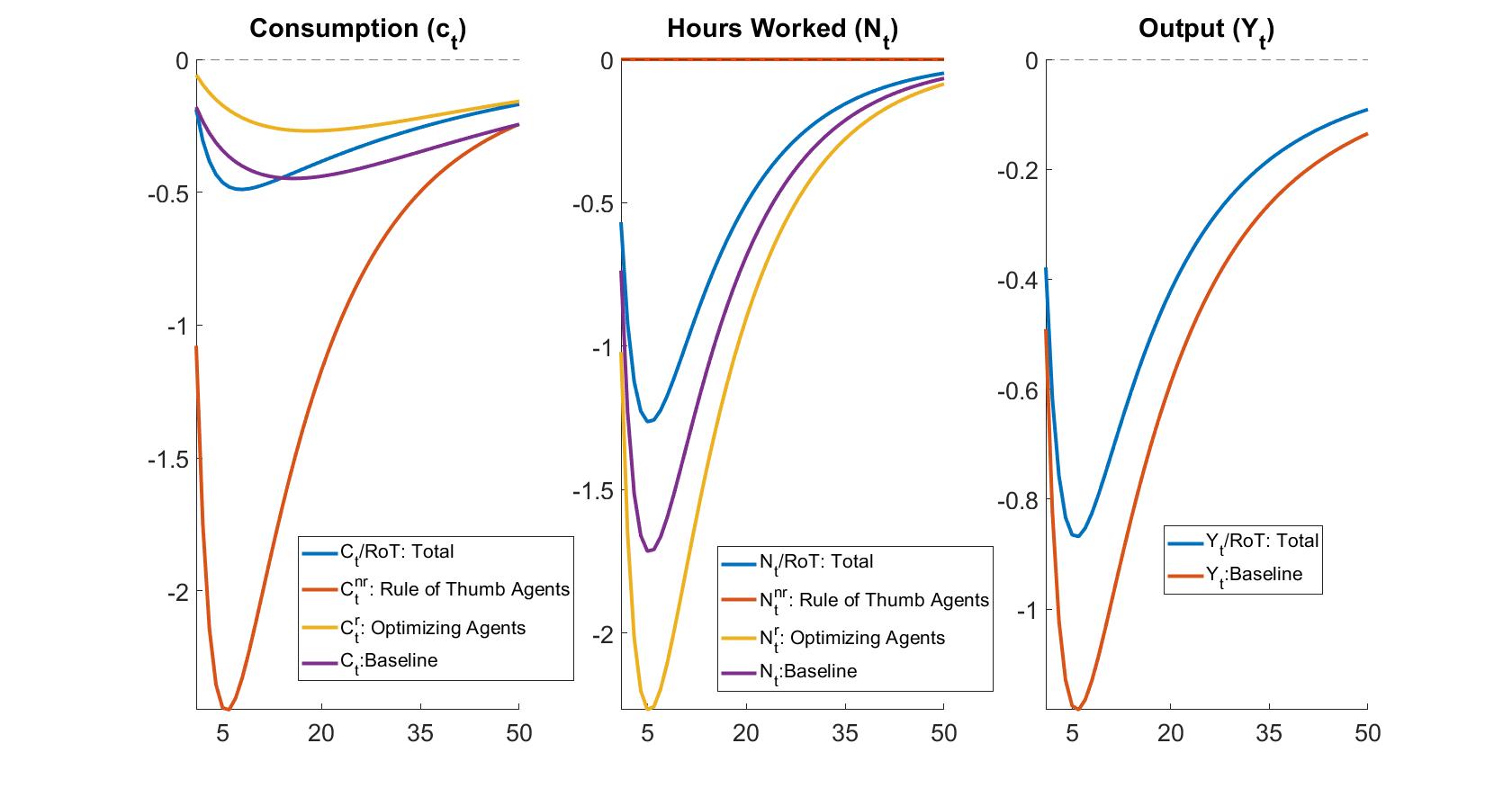} }
			\end{array}$	}
	\end{center} 
	\caption{Impulse responses for a 1 standard deviation shock to uncertainty $\eta_t^{*}$  in the extended  DSGE model with rule-of-thumb agents. An uncertainty shock to credit access leads to a disproportionately bigger decline in consumption for the rule-of-thumb agents. For the Ricardian households, the decline in consumption is smaller and decline in hours is bigger compared to baseline. Aggregate output falls by more in the baseline scenario.} \label{fig:Rot}
\end{figure}

In addition to Equation ($24$), aggregate consumption and hours now reflect the heterogeneity in households,
\begin{equation}
    C_t={\nu}C_t^{NR}+(1-\nu)C_t^{R}
\end{equation}and
\begin{equation}
    N_t={\nu}N_t^{NR}+(1-\nu)N_t^{R} \text{ .}
\end{equation}
The value of $\nu$ is now set to $\frac{1}{3}$ to understand  how the transmission mechanism is altered by the presence of these rule-of-thumb/non-Ricardian agents. The baseline model presented in sections 5.1 through 5.4 corresponds to $\nu=0$. The orange line in panel 1 in Figure ~\ref{fig:Rot} shows that the consumption of rule-of-thumb households fall disproportionately in response to an uncertainty shock in access to credit. Since hours supplied by the non-Ricardian households remain unchanged, hours supplied by the optimizing/Ricardian households contract more than the baseline (the yellow line in panel 2 of Figure~\ref{fig:Rot}). The blue line in panel 3 of Figure~\ref{fig:Rot} shows that in equilibrium, aggregate output falls by less compared to the baseline in orange in panel 2 of Figure~\ref{fig:Rot}. In this extension the smaller decline in output relative to the baseline is driven by a smaller decline in the aggregate supply in hours.


\section{Robustness Checks}
This section carries out some robustness checks of the results by considering alternative choice of priors for the SV parameters, taking a longer sample for the series $\{y_t\}$, and considering other measures of credit access variables. Table \ref{tab:Posterior-Mean-Estimates SV-Robustness} summarizes the results.
\paragraph{Alternative choice of prior:} We use the following tighter priors for the SV model 
$p\left(\phi_{y}\right)=p\left(\phi_{h}\right)\sim TN_{(0,1)}\left(0.9,0.05^{2}\right)$,
$p\left(\overline{\mu_h}\right)\sim N\left(0,10^{2}\right)$, $p\left(\tau\right)\sim TN_{(0,\infty)}\left(0.5,0.3^{2}\right)$,
and $p\left(\rho\right)\sim U\left(-1,1\right)$. The $TN_{(lo,up)}\left(c,d^{2}\right)$ denotes the univariate normal distribution with mean $c$ and standard deviation $d$ constrained to the interval $(lo,up)$.

\paragraph{Longer sample:} We use the period 1952Q1-2018Q4 to estimate the SV model.


\paragraph{Alternative definitions of credit:} Sections 2-5 examine the presence and impact of shocks to the second-moment using \textit{Total Credit to Private nonfinancial Sector} as a measure of credit growth. It captures the dynamics in shocks to aggregate credit access  as it is a broad measure of credit-availability. We now consider two other additional variables -- Total Consumer Credit Owned and Securitized, Flow,  ($x_{t}$ in Table
\ref{tab:Posterior-Mean-Estimates SV-Robustness}) and Total Consumer Credit Owned and Securitized, Percent Change at Annual Rate ($z_{t}$ in Table
\ref{tab:Posterior-Mean-Estimates SV-Robustness}).
\begin{table}[H]
	\caption{Posterior Mean Estimates (with 95\% credible intervals in brackets)
		of the SV. The ${y^{P}_{t}}$ and ${y^{L}_{t}}$ are the parameter estimates of the SV model with alternative choice of prior and with longer sample, respectively. \label{tab:Posterior-Mean-Estimates SV-Robustness}}	
	\centering{}%
	\begin{tabular}{cccccc}
		\hline 
		Parameters & $y_{t}$ -- Baseline & ${y^{P}_{t}}$ & ${y^{L}_{t}}$ & $x_{t}$ & $z_{t}$\tabularnewline
		\hline 
		\hline 
		$\overline{\mu_h}$ & $\underset{\left(-10.83,-9.61\right)}{-10.23}$ & $\underset{\left(-10.85,-9.67\right)}{-10.22}$ & $\underset{\left(-10.90,-9.06\right)}{-9.80}$ & $\underset{\left(-6.18,-4.58\right)}{-5.36}$ & $\underset{\left(-9.70,-6.56\right)}{-7.92}$\tabularnewline
		$\phi_{h}$ & $\underset{\left(0.75,0.98\right)}{0.91}$ & $\underset{\left(0.83,0.97\right)}{0.91}$ & $\underset{\left(0.93,0.99\right)}{0.97}$ & $\underset{\left(0.69,0.98\right)}{0.88}$ & $\underset{\left(0.76,0.99\right)}{0.91}$\tabularnewline
		$\phi_{y}$ & $\underset{\left(0.75,0.91\right)}{0.83}$ & $\underset{\left(0.78,0.91\right)}{0.84}$ & $\underset{\left(0.62,0.83\right)}{0.73}$ & $\underset{\left(0.88,0.99\right)}{0.94}$ & $\underset{\left(0.77,0.94\right)}{0.86}$\tabularnewline
		$\tau$ & $\underset{\left(0.12,0.53\right)}{0.27}$ & $\underset{\left(0.15,0.55\right)}{0.29}$ & $\underset{\left(0.08,0.28\right)}{0.16}$ & $\underset{\left(0.23,0.87\right)}{0.50}$ & $\underset{\left(0.26,0.84\right)}{0.52}$\tabularnewline
		$\rho$ & $\underset{\left(-0.16,0.95\right)}{0.50}$ & $\underset{\left(-0.21,0.92\right)}{0.45}$ & $\underset{\left(-0.06,0.76\right)}{0.42}$ & $\underset{\left(-0.59,0.18\right)}{-0.22}$ & $\underset{\left(-0.57,0.21\right)}{-0.19}$\tabularnewline
		\hline 
	\end{tabular}
\end{table}
Table \ref{tab:Posterior-Mean-Estimates SV-Robustness} shows that the parameter estimates of the SV model are very
similar between the three scenarios suggesting that the results are robust against
prior specification and different lengths of the datasets. Both the $x_{t}$ and
$z_{t}$ series have similar values of $\phi_{h}$, but bigger values of 
$\overline{\mu_h}$, $\tau$, $\phi_{y}$ and negative $\rho$ to the series $y_{t}$.

\section{Conclusion}
Our article presents a new stylized fact characterizing the time variation in the volatility of credit growth. It interprets this stochastic volatility in credit growth as uncertainty about access to credit. The results suggest that unforeseen changes in uncertainty about credit access can trigger a sharp slowdown in real activity. Furthermore, it shows that these effects are largely countercyclical, with bigger effects during recessions. A flexible-price model with collateral constraints is presented to provide intuition for the transmission mechanism, and the effects of an uncertainty shock to credit availability is examined in this framework. When the collateral constraint binds, an uncertainty shock about credit availability can generate a simultaneous decline in consumption, investment and output, a finding that previous work on uncertainty shocks without credit constraints has been unable to produce.

\newpage
\bibliographystyle{plainnat}
\bibliography{sample_pc2} 

\pagebreak
\newpage
\appendix
{\centering
\section*{Appendix}}
\section{The standard PMMH algorithm \label{alg:The-particle-marginal Mh}}

\begin{algorithm}[H]
\caption{The pseudo marginal Metropolis-Hastings\label{alg:standard}}

\begin{itemize}
\item Set the initial values of $\theta^{\left(0\right)}$ arbitrarily.
\item Sample $u\sim N\left(0,I\right)$, and run a particle filter to compute
an estimate $\widehat{p}\left(y_{1:T}|\theta,u\right)$ of the likelihood 
and to sample the initial $h_{1:T}^{\left(0\right)}$.
\item For each of the MCMC iterations, $i=1,...,M$,
\begin{itemize}
\item Sample $\theta^{'}$ from the proposal density $q\left(\theta^{'}|\theta\right)$
and $u^{'}\sim N\left(0,I\right)$
\item Run a particle filter to compute an estimate of likelihood $\widehat{p}\left(y_{1:T}|\theta^{'},u^{'}\right)$
and sample $h_{1:T}^{'}$.
\item With probability 
\begin{equation}
\min\left\{ 1,\frac{\widehat{p}\left(y_{1:T}|\theta^{'},u^{'}\right)p\left(\theta^{'}\right)}{\widehat{p}\left(y_{1:T}|\theta,u\right)p\left(\theta\right)}\frac{q\left(\theta|\theta^{'}\right)}{q\left(\theta^{'}|\theta\right)}\right\}\label{MH_pseudo};
\end{equation}
set $h_{1:T}^{\left(i\right)}=h_{1:T}^{'}$, $\widehat{p}\left(y_{1:T}|\theta,u\right)^{\left(i\right)}=\widehat{p}\left(y_{1:T}|\theta^{'},u^{'}\right)$,
and $\theta^{\left(i\right)}=\theta^{'}$; otherwise, set $h_{1:T}^{\left(i\right)}=h_{1:T}^{\left(i-1\right)}$,
$\widehat{p}\left(y_{1:T}|\theta,u\right)^{\left(i\right)}=\widehat{p}\left(y_{1:T}|\theta,u\right)^{\left(i-1\right)}$,
and $\theta^{\left(i\right)}=\theta^{\left(i-1\right)}$. 
\end{itemize}
\end{itemize}
\end{algorithm}

\section{The correlated PMMH algorithm \label{alg:The-correlated-particle-marginal Mh}}
\begin{algorithm}[H]
\caption{The correlated pseudo marginal Metropolis-Hastings (PMMH) \label{alg:correlated}}

\begin{itemize}
\item Set the initial values of $\theta^{\left(0\right)}$ arbitrarily.
\item Sample $u\sim N\left(0,I\right)$, and run a particle filter to compute
an estimate of likelihood $\widehat{p}\left(y_{1:T}|\theta,u\right)$
and to sample the initial $h_{1:T}^{\left(0\right)}$.
\item For each of the MCMC iterations, $i=1,...,M$,
\begin{itemize}
\item Sample $\theta^{'}$ from the proposal density $q\left(\theta^{'}|\theta\right)$. 
\item Sample $u^{*}\sim N\left(0,I\right)$, and set $u^{'}=\gamma u+\sqrt{1-\gamma^{2}}u^{*}$; $\gamma$ is the correlation between $u$ and $u^{'}$ and is
set close to 1. 
\item Run a particle filter to compute an estimate of likelihood $\widehat{p}\left(y_{1:T}|\theta^{'},u^{'}\right)$
and sample $h_{1:T}^{'}$.
\item With probability in Equation \eqref{MH_pseudo} 
set $h_{1:T}^{\left(i\right)}=h_{1:T}^{'}$, $\widehat{p}\left(y_{1:T}|\theta,u\right)^{\left(i\right)}=\widehat{p}\left(y_{1:T}|\theta^{'},u^{'}\right)$, $u^{(i)}=u^{'}$,
and $\theta^{\left(i\right)}=\theta^{'}$; otherwise, set $h_{1:T}^{\left(i\right)}=h_{1:T}^{\left(i-1\right)}$,
$\widehat{p}\left(y_{1:T}|\theta,u\right)^{\left(i\right)}=\widehat{p}\left(y_{1:T}|\theta,u\right)^{\left(i-1\right)}$,  $u^{(i)}=u^{(i-1)}$,
and $\theta^{\left(i\right)}=\theta^{\left(i-1\right)}$. 
\end{itemize}
\end{itemize}
\end{algorithm}

\section{The Correlated Particle Filter Algorithm \label{sec:The-Particle-Filter}}
This section discusses the correlated particle filter of \citet{Deligiannidis:2018}.
We use this particle filter algorithm to sequentially approximate the joint filtering 
densities \\ $\left\{ p\left(x_{t}|y_{1:t},\theta\right):t=1,...,T\right\} $
 using $N$ particles, i.e., weighted samples $\left\{ h_{t}^{1:N},\overline{w}_{t-1}^{1:N}\right\} $,
drawn from some proposal densities $m_{1}\left(h_{1}\right)$ and
$m_{t}\left(h_{t}|h_{t-1}\right)$ for $t\geq2$; see \citet{Andrieu:2010}
for detailed assumptions about the proposal densities. Let
\begin{equation}
w_{1}^{i}=\frac{p\left(y_{1}|h_{1}\right)p\left(h_{1}\right)}{m_{1}\left(h_{1}\right)}, w_{2}^{i}=\frac{p\left(y_{2}|h_{2},y_{1}\right)p\left(h_{2}|h_{1},y_{1}\right)}{m_{2}\left(h_{2}|h_{1}\right)}, w_{t}^{i}=\frac{p\left(y_{t}|h_{t}\right)p\left(h_{t}|h_{t-1}\right)}{m_{t}\left(h_{t}|h_{t-1}\right)},\;\textrm{for}\;t\geq3\;,
\end{equation}
and $\;\overline{w}_{t}^{i}=\frac{w_{t}^{i}}{\sum_{j=1}^{N}w_{t}^{j}}$.

Let $u$ be the pseudo-random vector used to obtain the unbiased
estimate of the likelihood; $u$ has two
components $u^{1:N}_{h,1:T}$ and $u^{1:N}_{A,1:T-1}$. 


Let $u_{h,t}^{i}$ be the vector
random variable used to generate the particles $h_{t}^{i}$ given
$\theta$ and $h_{t-1}^{a_{t-1}^{i}}$. We write, 
\begin{equation}
h_{1}^{i}=H\left(u_{h,1}^{i};\theta\right)\;\textrm{and}\;h_{t}^{i}=H\left(u_{h,t}^{i};\theta,h_{t-1}^{a_{t-1}^{i}}\right)\;\textrm{for}\;\;t\geq2,
\end{equation}
where $a_{t-1}^{i}$ is the ancestor index of $h_{t}^{i}$. Denote the distribution of $u_{h,t}^{i}$ as $\psi_{ht}(\cdot)$. 

For $t\geq2$, let $u_{A,t-1}$ be the vector of random variables used to generate the ancestor
indices $a_{t-1}^{1:N}$  using
the resampling scheme $\mathcal{M}\left(a_{t-1}^{1:N}|\overline{w}_{t-1}^{1:N},h_{t-1}^{1:N}\right)$ and define $\psi_{A,t-1}(\cdot)$ as the distribution of $u_{A,t-1}$. Common choices for $\psi_{ht}(\cdot)$ and $\psi_{A,t-1}(\cdot)$ are iid $U(0,1)$ or iid $N(0,1)$ random variables.
The particle filter provides the unbiased estimate of the likelihood
\begin{equation}
\widehat{p}\left(y|\theta,u\right)=\prod_{t=1}^{T}\left(\frac{1}{N}\sum_{i=1}^{N}w_{t}^{i}\right).
\end{equation}
For the correlated PMMH to work efficiently, it is necessary that the logs of the likelihood estimates
$\widehat{p}\left(y|\theta,u\right)$ and $\widehat{p}\left(y|\theta^{'},u^{'}\right)$
are highly correlated. The resampling steps in the particle filter
introduce discontinuities even when $\theta$ and $\theta^{'}$ are only
slightly different, where $\theta$ is the current value and $\theta^{'}$ is the proposed value of the parameters. The discontinuity problem is solved by first sorting the
particles from the smallest to largest before resampling them. 
Algorithm \ref{alg:The-correlated particle filter} gives the correlated particle filter algorithm. At time $t=1$, step 1 generates
\begin{equation}
h_{1}^{i}=\sqrt{{\tau^{2}}/{(1-\phi^{2})}}u_{h,1}^{i}+\overline{\mu_h},
\end{equation}
for $i=1,...,N$. Steps 2 and 3 compute unnormalised and normalised weights for all particles, respectively.

At time $t>1$, Step 1 sorts the particles from smallest to largest and obtains the sorted particles and weights. Step 2 in Algorithm \ref{alg:Multinomial-Resampling-Algorithm} resamples the particles using multinomial resampling to obtain the ancestor index $A_{1:T-1}^{1:N}$ in the original order of the particles in Step 3. Step 4 generates 
\begin{equation}
h_{t}^{i}=\overline{\mu_h}+\phi\left(h_{t-1}^{a_{t-1}^{i}}-\overline{\mu_h}\right)+\rho\tau\exp\left(-\frac{h_{t-1}^{a_{t-1}^{i}}}{2}\right)\left(y_{t-1}-\phi_{y}y_{t-2}\right)+\sqrt{\tau^{2}\left(1-\rho^{2}\right)}u_{h,t}^{i},
\end{equation}
for $i=1,...,N$; it then compute the unnormalised and normalised weights of all particles.



\begin{algorithm}[H]
\caption{The Correlated particle filter (CPF) algorithm \label{alg:The-correlated particle filter} }

Input: $u^{1:N}_{h,1:T}$, $u^{1:N}_{A,1:T-1}$, and $N$

Output: $h^{1:N}_{1:T}$, $A^{1:N}_{1:T-1}$, and $\overline{w}^{1:N}_{1:T}$

For $t=1$,
\begin{enumerate}
\item Set $h_{1}^{i}=H\left(u_{h,1}^{i};\theta\right)$ for $i=1,...,N$.
\item Compute the unnormalised weights $w_{1}^{i}$,
for $i=1,...,N$.
\item Compute the normalised weights $\overline{w}_{1}^{i}$
for $i=1,...N$.
\end{enumerate}
For $t\geq2$
\begin{enumerate}
\item Sort the particles $h_{t-1}^{i}$ from the smallest to largest and obtain the sorted indices $\zeta_{i}$ for $i=1,...,N$, and the sorted particles and weights $\wt h_{t-1}^i = h_{t-1}^{\zeta_i} $ and $\wt {\ov w}^i_{t-1} = \ov w_{t-1}^{\zeta_i}$, for $i=1, \dots, N$.

\item Obtain the ancestor indices based on the sorted particles $\wt a_{t-1}^{1:N}$ using the multinomial resampling in Algorithm~\ref{alg:Multinomial-Resampling-Algorithm}.

\item Obtain the ancestor indices based on the original order of the particles $a_{t-1}^{i}$ for $i=1,...,N$.

\item Set $h_{t}^i =H\left(u_{ht}^{i}; \theta, h_{t-1}^{a_{t-1}^i} \right)$,
for $i=1,...,N$.
\item Compute the unnormalised weights $w_{t}^{i}$,
for $i=1,...,N$
\item Compute the normalised weights $\overline{w}_{t}^{i}$
for $i=1,...,N$
\end{enumerate}
\end{algorithm}

\begin{algorithm}[H]
\caption{Multinomial Resampling Algorithm \label{alg:Multinomial-Resampling-Algorithm}}

Input: $u_{Amt-1}$, sorted particles $\widetilde{h}_{t-1}^{1:N}$, and sorted weights $\widetilde{\overline{w}}_{t-1}^{1:N}$ (see Algorithm \ref{alg:The-correlated particle filter})

Output: $\widetilde{a}_{t-1}^{1:N}$ 
\begin{enumerate}
\item Compute the cumulative weights 
\[
\widehat{F}_{t-1}^{N}\left(j\right)=\sum_{i=1}^{j}\widetilde{\overline{w}}_{t-1}^{i}.
\]
based on the sorted particles $\left\{ \widetilde{h}_{t-1}^{1:N},\widetilde{\overline{w}}_{t-1}^{1:N}\right\} $
\item Set $\widetilde{a}_{t-1}^{i}=\underset{j}{\min} \quad \widehat{F}_{t-1}^{N}\left(j\right)\geq u_{A,t-1}^{i}$
for $i=1,...N$; For $i=1,...,N$ $\widetilde{a}_{t-1}^{i}$ 
is the ancestor index based on the sorted particles.
\end{enumerate}
\end{algorithm}

\end{document}